\documentstyle[12pt]{article}
\newcommand{\bm}[1]{\mbox{\boldmath $#1$}}
\def\be{\begin{equation}}
\def\ee{\end{equation}}
\def\bea{\begin{eqnarray}}
\def\eea{\end{eqnarray}}
\def\b*{\begin{eqnarray*}}
\def\e*{\end{eqnarray*}}
\def\N{\hfill \rule{2.5mm}{2.5mm}}
\def\R{{\rm I\!R}}
\def\U{\Upsilon}
\def\S{\Sigma}
\def\O{\Omega}
\def\DP{{\cal DP}}
\def\SE{{\cal SE}}
\def\SS{{\cal SS}}
\def\NS{{\cal NS}}

\def\P{{\it Proof:} \hspace{3mm}}

\newtheorem{defi}{Definition}[section]
\newtheorem{theo}{Theorem}[section]
\newtheorem{coro}{Corollary}[section]
\newtheorem{prop}{Proposition}[section]
\newtheorem{lem}{Lemma}[section]
\newtheorem{pr}{Property}[section]

\title{Null cone preserving maps, causal tensors and algebraic
Rainich theory} 
\author{G\"{o}ran Bergqvist$^{1,2}$, Jos\'{e} M. M. Senovilla$^2$ \\
$^1$ Department of Mathematics, M{\"a}lardalen University,\\
721 23 V{\"a}ster{\aa}s, Sweden \\
E-mail: gbt@mdh.se \\
$^2$ Departamento de F\'{\i}sica Te\'orica, Universidad del Pa\'{\i}s Vasco,\\
Apartado 644, 48080 Bilbao, Spain \\
E-mail: wtpmasej@lg.ehu.es}
\begin{document}           

\maketitle                 


\abstract{
A rank-$n$ tensor on a Lorentzian manifold whose contraction with $n$ arbitrary 
causal future directed vectors is non-negative is said to have the dominant 
property.
These tensors, up to sign, are called {\it causal} 
tensors, and we determine their general mathematical properties in arbitrary
dimension $N$. Then, we prove that rank-2 tensors 
which map the null cone on itself are causal tensors. Previously it has been 
shown that, to any tensor field $A$ on a
Lorentzian manifold there is a corresponding ``superenergy'' tensor field
$T\{A\}$ (defined as a quadratic sum over all Hodge duals of $A$) 
which always has the dominant property. 
Here we prove that, conversely, any symmetric rank-2 tensor with the 
dominant property can be written in a canonical way as a sum of $N$ 
superenergy tensors of {\it simple} forms.
We show that the square of any rank-2 superenergy tensor is proportional
to the metric in dimension $N\le 4$, and that the square of the superenergy 
tensor of any simple form is proportional to the metric in arbitrary dimension.
Conversely, we prove in arbitrary dimension that any symmetric rank-2 
tensor $T$ whose square is proportional to the metric must be a causal 
tensor and, up to sign, the superenergy of a simple $p$-form, and that the 
trace of $T$ determines the rank $p$ of the form. This generalises, both 
with respect to the dimension $N$ and the rank $p$, the
classical algebraic Rainich conditions, which are necessary and sufficient 
conditions for a metric to originate algebraically in some physical field.
Furthermore, it has the important geometric interpretation that the set of
superenergy tensors of simple forms is precisely the set of tensors which
leave the null cone invariant and preserve its time orientation. 
It also means that all involutory Lorentz transformations can be represented
as superenergy tensors of simple forms, and that any rank-2 superenergy tensor
is the sum of at most $N$ conformally involutory Lorentz transformations.
Non-symmetric null cone preserving maps are shown to have a symmetric part
with the dominant property and are classified according to the null
eigenvectors of the skew-symmetric part. We therefore obtain a complete
classification of all conformal Lorentz transformations and singular null
cone preserving maps on any Lorentzian manifold of any dimension.}

\vspace{1cm}

\newpage

\section{Introduction}

The Bel-Robinson tensor \cite{B,Bel2}, a rank-4 tensor constructed from the Weyl
curvature tensor and its dual (it has only one dual in {\it four} dimensions),
was until some ten years ago not a
widely known tensor outside part of the general relativity community.
That it has the dominant property ---the contraction with any four causal
future directed vectors is non-negative--- was certainly known \cite{PR}, and
many relations to gravitational energy were found (see e.g. \cite{S2} and
references therein). Its precise physical
meaning was, and still is, however not clear, and it is possible that
no fundamental physical interpretation can be given. Thus, interest
in the Bel-Robinson tensor was limited. This all changed with the
work of Christodoulou and Klainerman on the global non-linear stability
of Minkowski spacetime \cite{CK}, (Bel-Robinson estimates were in fact
previously considered in the works by
Friedrich on hyperbolic formulations of the field equations, see \cite{F}.)
It became clear that the Bel-Robinson tensor
is {\it mathematically} a very useful quantity, its positivity (the
dominant property) and divergence properties being the main reasons.
Today, the tensor is established as a key ingredient in many mathematical
studies of Einstein's vacuum equations, see e.g. \cite{KN,Ren} and
references therein.

Considering this rise of interest in the Bel-Robinson tensor, it is remarkable
that the Bel tensor seems virtually unknown. This is the full Riemann
curvature tensor analogue of the Bel-Robinson tensor, so it is
constucted from the Riemann tensor and its duals \cite{Bel,S2}. A fundamental fact
is that also the Bel tensor has the dominant property \cite{B2,Bo,S2}, and it is
essentially the only tensor with this property one can construct
from the Riemann tensor. Its divergence can often be controlled if
some suitable field equations for the matter are given, and it should therefore
be the natural candidate to replace the Bel-Robinson tensor if the
full Einstein's equations are studied.

More recently, it was discovered
that this way of constructing a tensor with the dominant property from a
given tensor and its duals is universal \cite{S1,S2}. Given {\it any} tensor field $A$
on a Lorentzian manifold of arbitrary dimension, one can always in an essentially
unique way construct from $A$ a corresponding tensor
$T\{ A\}$ with the dominant property \cite{B2,PP,S2}. It is perhaps unfortunate that, by
historical reasons, $T\{ A\}$ has become to be known as the {\it superenergy}
tensor of $A$, as this terminology may have prevented attention from
those studying differential equations on curved manifolds. Superenergy
tensors provide a very natural and geometric way to define
norms (including Sobolev norms) and inner products (corresponding to
the positive norms) on Lorentzian manifolds. Like the Bel-Robinson tensor,
there is no need of a physical interpretation of $T\{ A\}$ for it to be
mathematically useful.

A first example of how the general superenergy tensors can be used in this
sense was given in \cite{BS}, where causal
propagation of fields on Lorenztian manifolds was studied generalising
techniques from \cite{HE,BoS}. Note that for energy-momentum tensors
(symmetric rank-2 tensors) the dominant property, first introduced
in \cite{Ple}, is usually called
the dominant energy condition \cite{HE}. Such tensors map the future cone
on itself, something we refer to as a causal map or {\it causal tensor}.
Superenergy tensors have also been used to construct new conserved
quantities \cite{S2,S3}, and to study the propagation
of shock-waves \cite{S2}.

\bigskip

In this paper we develop the mathematical structure of tensors having the
dominant property and prove some new basic results about superenergy
tensors. We prove that the product $T_{ac}T_{b}{}^{c}$
of the superenergy tensor $T_{ab}\{ A\}$ of a {\it{simple}}
form $A$ is always proportional to the metric. This is further shown to be
true for any arbitrary rank-2 superenergy tensor in dimensions $N\leq 4$.

While any superenergy tensor has the dominant property, we prove that
any symmetric rank-2 tensor with the dominant property can be written
as a sum of $N$ superenergy tensors of simple forms in a canonical way.
We also present some geometrical interpretation of these forms and emphasize
that, in this sense, superenergy tensors of simple forms are the basic
building blocks of positive or causal quantities.

The classical Rainich conditions \cite{R,MW}, sometimes referred to
as RMW (Rainich-Misner-Wheeler) theory or already unified theory, are
necessary and sufficient conditions in 4 dimensions for an
energy-momentum tensor to originate in a Maxwell field. One may also
express this as saying that they are conditions on a metric, which
then via the Ricci tensor and Einstein's equations give the
energy-momentum tensor. The algebraic Rainich conditions as stated
in \cite{PR} are that the energy-momentum tensor is trace-free,
satisfies the dominant energy condition, and has a square
that is proportional to the metric. The Rainich conditions have also
been generalised to cover some other physical situations
(e.g. \cite{CF,Ku,Pen1,Pen2,Per}).

Here, we prove a much more general result, namely that in $N$ dimensions
any symmetric rank-2 tensor with a square proportional to the
metric must be the superenergy tensor of a simple $p$-form. We prove that
the trace can only have certain discrete values related to the
rank $p$ of this form.
This result, being an \underline{equivalence}, has an important geometrical
interpretation. It says that on any Lorentzian manifold of any
dimension, the set of superenergy tensors
of simple forms is precisely the set of tensors which leave
the null cone invariant and preserve its time orientation. This also leads
to an extended algebraic Rainich theory which includes the previously
known results as special cases.
It also has the interesting implication that all symmetric (i.e. involutory)
Lorentz transformations can be represented as superenergy tensors
of simple forms. Furthermore, the combination with the previous results
proves that any superenergy tensor is the sum of at most $N$
conformally involutory Lorentz transformations.
We also study non-symmetric null cone
preserving maps, which are proven to have a symmetric part with the
dominant property, and classify them according to the null eigenvectors of its
skew-symmetric part. All this provides a complete classification of
all conformal Lorentz transformations as well as the singular null
cone preserving maps in any Lorentzian manifold of arbitrary dimension.

\bigskip

In our notation we sometimes use indices on the tensors. These indices may
be considered as abstract indices in the sense of Penrose and
Rindler \cite{PR}, and it is clear that all results are geometric
and independent of any basis or coordinate system. We will also use
the standard arrows for vectors and boldface characters for 1-forms.
The tensor and exterior
products are denoted by $\otimes$ and $\wedge$ respectively.
As usual, (square) round brackets enclosing any set of indices indicate
(anti) symmetrization.
Equalities by definition are denoted by $\equiv$. The symbol
\rule{2.5mm}{2.5mm} is used to mark the end of proofs.
We shall use the signature $+,-,\dots ,-$ of the metric.
Note that this is the opposite of \cite{S2}.

The outline of the paper is as follows. In section 2 we develop some general
mathematical properties of tensors with the dominant property and in section 3
we recall the definition of superenergy tensors and prove certain new
results for
superenergy tensors of simple forms. We also extend the previous definition
to $N$-forms in $N$ dimensions and motivate why their superenergy is
essentially the metric. Various properties of null cone
preserving maps, their relation to superenergy tensors,
and how these are used to construct any tensor satisfying the dominant
energy condition are described in section 4. The classification of
the (conformally) non-involutory Lorentz transformation is then
given in section 5 while that of the (conformally) involutory ones and
the generalised algebraic Rainich conditions with their
important geometrical consequences are presented in section 6.

\section{The dominant property: causal tensors}
We assume that we work on an $N$-dimensional manifold $V_N$ endowed with
a Lorentzian metric $g_{ab}$ and that a time-orientation has been chosen.
Most of our considerations are algebraic and
are implicitly assumed to hold in a point $x\in V_N$; of course, they can
be straightforwardly translated to tensor fields.
We begin by giving the basic definition.
\begin{defi}
A tensor $T_{a_{1}...a_{r}}$ is said to
have the \underline{dominant property} if
$$T_{a_{1}...a_{r}}u_{1}^{a_{1}}...u_{r}^{a_{r}}\geq 0$$ for any set
$u_{1}^{a_{1}}$,..., $u_{r}^{a_{r}}$ of causal future-pointing vectors.
The set of tensors with the dominant property will be denoted by $\DP$.
By $-\DP$ we mean the set of tensors $T_{a_{1}...a_{r}}$ such that
$-T_{a_{1}...a_{r}}\in \DP$.
\label{DP}
\end{defi}
We will see below, in Properties \ref{strict} and \ref{null} respectively,
that the definition of $\DP$ implies in fact that the strict inequality holds
if the future-pointing $u_{1}^{a_{1}}\dots u_{r}^{a_{r}}$ are all timelike,
and
that the use of only null vectors $u_{1}^{a_{1}}\dots u_{r}^{a_{r}}$ is
also enough.

By a natural extension, the non-negative real numbers are also considered
to have the dominant property: $\R^+ \subset \DP$.
Rank-1 tensors with the dominant property are simply the future-pointing
causal vectors (while those in $-\DP$ are the past-directed ones). For
rank-2 tensors, the dominant property was introduced by Pleba\'{n}ski
\cite{Ple} in General Relativity and is usually called the dominant energy
condition \cite{HE} because it is a requirement for physically
acceptable energy-momentum tensors. The elements of $\DP$ could thus be termed
as ``future tensors'', and those of $\DP \cup -\DP$ will be called
``causal tensors''.
As in the case of past- and future-pointing vectors, any statement
concerning $\DP$ has its counterpart concerning $-\DP$, and they will be
taken as obvious unless otherwise stated.

The basic properties of tensors in the class $\DP$ are given in what follows.
First of all, the class is closed under linear combinations with
non-negative coefficients as well as under tensor products \cite{S2}.
\begin{pr}
If $T_{a_{1}...a_{r}}^{(i)}\in \DP$ and $\alpha _{i}\in \R^+$ ($i=1,...,n$)
then $\sum\limits_{i=1}^{n}\alpha _{i}T^{(i)}_{a_{1}...a_{r}}\in\DP$.
Moreover, if
$T_{a_{1}...a_{r}}^{(1)}$ , $T_{a_{1}...a_{s}}^{(2)}\in \DP$ then
$\left(T^{(1)}\otimes T^{(2)}\right)_{a_1\dots a_{r+s}} \in \DP$.
\label{pr:alg}
\end{pr}
\P This is an immediate consequence of the definition of $\DP$. \N

Given any tensor $T_{a_{1}...a_{r}}\in \DP$, one can immediately
construct many other tensors in $\DP$ by simply permuting the indices,
as is obvious from Definition \ref{DP}. Then, we also have
(see Section 5 in \cite{S2})
\begin{lem}
If $T_{a_{1}...a_{r}}\in \DP$, then for any set of non-negative
constants $c_{\sigma}$ the family of tensors
$\sum\limits_{\sigma} c_{\sigma}T_{a_{\sigma (1)}\dots a_{\sigma (r)}}$
belongs to $\DP$ where the sum is over all permutations
$\sigma (1),\dots ,\sigma(r)$ of $(1,\dots ,r)$.
In particular, any symmetric part of $T_{a_{1}...a_{r}}$ is in $\DP$.
\label{lem:perm}
\end{lem}
\P Given that $T_{a_{\sigma (1)}\dots a_{\sigma (r)}}\in \DP$
for any permutation $\sigma (1),\dots ,\sigma(r)$ the first part follows
from Property \ref{pr:alg}. Since any symmetric part is in fact a linear
combination of such terms with particular positive coefficients $c_{\sigma}$
the Lemma is proven. \N

It must be remarked that, sometimes, linear combinations
$\sum\limits_{\sigma} c_{\sigma}T_{a_{\sigma (1)}\dots a_{\sigma (r)}}$
with some negative coefficients $c_{\sigma}$ may also be in $\DP$.
On the other hand, we also have
\begin{lem}
If $T_{a_{1}...a_{r}}\neq 0$ is antisymmetric in any pair of indices, then
$T_{a_{1}...a_{r}}$ cannot be in $\DP\cup -\DP$.
\label{lem:skew}
\end{lem}
\P Assume, for instance, that $T_{a_{1}a_2...a_{r}}=-T_{a_{2}a_1...a_{r}}$.
Then, for any future-pointing $u_1^{a_1},u_2^{a_2},\dots ,u_r^{a_r}$,
the scalars $T_{a_{1}a_2...a_{r}}u_1^{a_1}u_2^{a_2}\dots u_r^{a_r}$ and
$T_{a_{1}a_2...a_{r}}u_2^{a_1}u_1^{a_2}\dots u_r^{a_r}$ have opposite
signs. This implies that a constant sign cannot be maintained.\N
\begin{pr}
$T_{a_{1}...a_{r}}\in \DP \Longleftrightarrow
u^{a_r}T_{a_{1}a_2...a_{r}}\in\DP$
for all future-pointing vectors $\vec{u}$.
\label{pr:dp}
\end{pr}
\P Again this is trivial from Definition \ref{DP}. \N

Of course, this can be equally proven for the contraction of $\vec{u}$ with
any index of $T_{a_{1}...a_{r}}$.
The previous property can be generalized to show that the class $\DP$
is also closed under tensor products with {\em one} contraction applied. To
that
end, we introduce the following products for any two tensors
$T_{a_{1}...a_{r}}^{(1)}$ and $T_{a_{1}...a_{s}}^{(2)}$:
$$
(T^{(1)}\, {}_i\!\times_j\, T^{(2)})_{a_1\dots\dots a_{r+s-2}}\equiv
T^{(1)}_{a_{1}...a_{i-1}ba_{i}...a_{r-1}}
T^{(2)}_{a_{r}...a_{r+j-2}}{}^{b}{}_{a_{r+j-1}...a_{r+s-2}}
$$
where the contraction is taken with the $i^{th}$ index of the first tensor and
the $j^{th}$ of the second. There are of course many different products
${}_i\!\times_j$ depending on where the contraction is made.
\begin{lem} For all $i=1,\dots ,r$ and all $j=1,\dots ,s$,
if $T_{a_{1}...a_{r}},t_{a_{1}...a_{s}}\in \DP$ or if
$T_{a_{1}...a_{r}},t_{a_{1}...a_{s}}\in -\DP$, then
$(T\, {}_i\!\times_j\, t)_{a_1\dots\dots a_{r+s-2}}\in \DP$; and if
$T_{a_{1}...a_{r}}\in DP$ and $t_{a_{1}...a_{s}}\in -\DP$ then
$(T\, {}_i\!\times_j\, t)_{a_1\dots\dots a_{r+s-2}}\in -\DP$ and
$(t\,\, {}_j\!\times_i T)_{a_1\dots\dots a_{r+s-2}}\in -\DP$.
\label{lem:cont}
\end{lem}
\P If $T_{a_{1}...a_{r}},t_{a_{1}...a_{s}}\in \DP$, or if they are in $-DP$,
then, by Property \ref{pr:dp},
\bea v_{b}&\equiv& T_{a_{1}...a_{i-1}ba_{i}...a_{r-1}}
u_{1}^{a_{1}}...u_{i-1}^{a_{i-1}}u_{i}^{a_{i}}...u_{r-1}^{a_{r-1}}\, ,
\label{v}\\
w_{b}&\equiv& t_{a_{r}...a_{r+j-2}ba_{r+j-1}\dots a_{r+s-2}}
u_{r}^{a_{r}}...u_{r+j-2}^{a_{r+j-2}}
u_{r+j-1}^{a_{r+j-1}}...u_{r+s-2}^{a_{r+s-2}}\nonumber
\eea
are causal vectors with the same time orientation for any set
$u_{1}^{a_{1}}\dots u_{r+s-2}^{a_{r+s-2}}$ of future-pointing vectors.
Hence
\b*
(T\, {}_i\!\times_j\, t)_{a_1\dots\dots a_{r+s-2}}
u_{1}^{a_{1}}...u_{r+s-2}^{a_{r+s-2}}=v_{b}w^{b}\geq 0 \hspace{2cm}
\e*
and the first result follows. The other is similar. \N
\begin{coro}
If for some $i,j$,
$0\neq (T\, {}_i\!\times_j T)_{a_1\dots\dots a_{2r-2}}\in -\DP$,
then $T_{a_{1}...a_{r}}$ cannot be in $\DP \cup -\DP$.
\label{cor:+}
\end{coro}
\P For if $T_{a_{1}...a_{r}}$ were in either $\DP$ or $-\DP$, by Lemma
\ref{lem:cont} $(T\, {}_i\!\times_j T)_{a_1\dots\dots a_{2r-2}}$ should be
in $\DP$. Notice that $\DP \cap -\DP =\{0\}$.\N
\begin{coro} $T_{a_{1}...a_{r}}\in \DP \Longleftrightarrow
(T\, {}_i\!\times_j\, t)_{a_1\dots\dots a_{r+s-2}}\in \DP$
for all $t_{a_{1}...a_{s}}\in \DP$.
\label{cor:cont}
\end{coro}
\P The implication from left to right is in Lemma \ref{lem:cont}.
The converse can be proved by taking in particular $s=1$ and using
Property \ref{pr:dp}. \N

\noindent
Corollary \ref{cor:cont} is the evident generalization of the well-known fact
that a causal vector $\vec{u}$ is future-pointing if and only if
$u_cv^c\in \R^+$ for all future-pointing vectors $\vec{v}$.

The concept of ``positivity'' does not capture all that is behind the
definition
of $\DP$, and the terminology dominant property (or dominant energy condition)
is preferable because the pure time component dominates any other component in
orthogonal bases.
\begin{lem}
$T_{a_{1}...a_{r}}\in \DP \ \Longleftrightarrow \
T_{0...0}\geq \left| T_{\alpha _{1}...\alpha_{r}}\right| $ for all
$\alpha _{1},...,\alpha _{r}\in \left\{0,1,...,N-1\right\} $, where
$T_{\alpha _{1}...\alpha _{r}}$ are the components of $T_{a_{1}...a_{r}}$
with respect to {\em any} orthonormal basis
$\left\{\vec{e}_{0},\vec{e}_{1},...,\vec{e}_{N-1}\right\}$ with
a future-pointing timelike $\vec{e}_{0}$.
\label{lem:dp}
\end{lem}
\P See Lemma 4.1 in \cite{S2}.\N
\begin{coro}
If $T_{a_{1}...a_{r}}\in \DP$ and $T_{a_{1}...a_{r}}u^{a_1}\dots u^{a_r}=0$
for a timelike vector $\vec{u}$, then $T_{a_{1}...a_{r}}=0$.
\label{cor:dp}
\end{coro}
\P By choosing the sign $\epsilon$ we have that $\vec{e}_{0}\equiv
\epsilon \, \vec{u}/(u_a u^a)$ is unit and future-pointing. Thus, by Lemma
\ref{lem:dp}, all components of $T_{a_{1}...a_{r}}$ vanish in any
orthonormal basis including $\vec{e}_{0}$, which means that
$T_{a_{1}...a_{r}}$
is the zero tensor.\N
\begin{coro}
If $T_{a_{1}...a_{r}}\in \DP$ and $T_{a_{1}...a_{r}}u^{a_r}=0$
for a timelike vector $\vec{u}$, then $T_{a_{1}...a_{r}}=0$.\N
\label{cor:dp'}
\end{coro}

Definition \ref{DP} involves all causal future-pointing vectors, but in fact
the class $\DP$ can be equally characterized by using timelike vectors
exclusively, or also only null vectors. Concerning the timelike case we have:
\begin{pr} A tensor $T_{a_{1}...a_{r}}\ne 0$ is in
$\DP \ \Longleftrightarrow \
T_{a_{1}...a_{r}}u_{1}^{a_{1}}...u_{r}^{a_{r}}> 0$
for any set $u_{1}^{a_{1}},...,u_{r}^{a_{r}}$ of timelike future-pointing
vectors.
\label{strict}
\end{pr}
\P The implication from right to left follows by continuity.
Conversely, first for rank 1, $T_{a_{1}}u_{1}^{a_{1}}=0$ for
$T_{a_{1}}\in \DP$ would imply that $T_{a_{1}}$ and $u_{1}^{a_{1}}$
are parallel and null as this is the only way two causal vectors
can be orthogonal. Thus, if $u_{1}^{a_{1}}$ is timelike and
$T_{a_{1}}\ne 0$ then $T_{a_{1}}u_{1}^{a_{1}}>0$. Suppose now that the
property has been proved for rank-$(r-1)$ tensors and
define $\tau_{a_{1}...a_{r-1}}\equiv T_{a_{1}...a_{r}}u_{r}^{a_{r}}$.
By Property \ref{pr:dp} $\tau_{a_{1}...a_{r-1}}\in \DP$ and hence,
if $T_{a_{1}...a_{r}}u_{1}^{a_{1}}...u_{r}^{a_{r}}=0$ then
$\tau_{a_{1}...a_{r-1}}u_{1}^{a_{1}}...u_{r-1}^{a_{r-1}}=0$ which in turn,
by Corollary \ref{cor:dp}, would
imply that $\tau_{a_{1}...a_{r-1}}=T_{a_{1}...a_{r}}u_{r}^{a_{r}}=0$.
But then Corollary \ref{cor:dp'} would lead to $T_{a_{1}...a_{r}}=0$.
Thus the result follows by induction on $r$. \N

In order to give the characterization with null vectors we first need a
basic result stating that future-pointing null vectors are the basic
``building blocks'' of all future-pointing vectors, i.e. rank-1 tensors
in $\DP$. In section 4 we shall generalize this by identifying the analogous
building blocks of rank-2 tensors in $\DP$.
\begin{lem} Given a future-pointing timelike vector $\vec{u}$ and a
future-pointing null vector $\vec{k}$, there is another future-pointing
null vector $\vec{n}$ such that $\vec{u}=c\vec{k}+\vec{n}$ where
$c=u^{a}u_{a}/(2u^{a}k_{a})>0$.
\label{lem:bb1}
\end{lem}
\P See, e.g., \cite{B1}.\N

\noindent
We can now show that in order to check that a
tensor is in $\DP$ it is sufficient to check it for null vectors.
This is very helpful because obviously it is easier to
work with null vectors exclusively rather than with both
null and timelike vectors.
\begin{pr} $T_{a_{1}...a_{r}}\in \DP \Longleftrightarrow
T_{a_{1}...a_{r}}k_{1}^{a_{1}}...k_{r}^{a_{r}}\geq 0$ for any set
$k_{1}^{a_{1}}$,..., $k_{r}^{a_{r}}$ of
future-pointing {\em null} vectors.
\label{null}
\end{pr}
\P By Lemma \ref{lem:bb1}, $T_{a_{1}...a_{r}}u_{1}^{a_{1}}...u_{r}^{a_{r}}$
with $s\leq r$ timelike vectors can be written as a sum with positive
coefficients of $2^{s}$ terms of
the type $T_{a_{1}...a_{r}}k_{1}^{a_{1}}...k_{r}^{a_{r}}$ involving null
vectors only and the result follows immediately. \N

Now we can prove a partial but important converse of the Lemma \ref{lem:cont}.
\begin{prop}
$0\neq \left(T{}_i\!\times_i\! T\right)_{a_1\dots a_{2r-2}}\in \DP$
for some $i=1,\dots ,r \, \Longrightarrow
T_{a_1\dots a_r}\in \DP \cup -\DP$.
\label{prop:square}
\end{prop}
\P  For any
set of {\it timelike} future pointing vectors
$u_{1}^{a_{1}},...,u_{r-1}^{a_{r-1}}$
define $v_b$ as in (\ref{v}) of Lemma \ref{lem:cont}. Now,
if $\left(T{}_i\!\times_i\! T\right)_{a_1\dots a_{2r-2}}\in \DP$
then $v_b v^b\geq 0$ which implies that $v_b$ is causal, either future- or
past-pointing. To see that the time orientation of these $v_b$ is consistent
take any other arbitrary set of timelike future-pointing vectors
$\tilde{u}_{1}^{a_{1}},...,\tilde{u}_{r-1}^{a_{r-1}}$ and define
$\tilde{v}_b$ analogously to (\ref{v}). As
$\left(T{}_i\!\times_i\! T\right)_{a_1\dots a_{2r-2}}\in \DP$ and is not
zero by assumption, using Property \ref{strict} we have that
$v_b\tilde{v}^b> 0$, so that $v_b$ and $\tilde{v}_b$ have the same
time orientation. As all the vectors $u_{1}^{a_{1}},...,u_{r-1}^{a_{r-1}}$
and $\tilde{u}_{1}^{a_{1}},...,\tilde{u}_{r-1}^{a_{r-1}}$ are arbitrary
and future pointing, this means that $T_{a_{1}...a_{r}}$ or
$-T_{a_{1}...a_{r}}$ is in $\DP$. \N
\begin{coro}
$0\neq \left(T{}_i\!\times_i\! T\right)_{a_1\dots a_{2r-2}}\in \DP$
for some $i=1,\dots ,r \Longrightarrow
\left(T{}_i\!\times_j\! T\right)_{a_1\dots a_{2r-2}} \in \DP$
for {\em all} $i,j=1,\dots ,r$.
\label{cor:squares}
\end{coro}
\P If $0\neq \left(T{}_i\!\times_i\! T\right)_{a_1\dots a_{2r-2}}\in \DP$
for some $i=1,\dots ,r$ then by Proposition \ref{prop:square}
$\epsilon T_{a_1\dots a_r}\in \DP$ with $\epsilon^2=1$, but then by
Lemma \ref{lem:cont}
$\left(T{}_i\!\times_j\! T\right)_{a_1\dots a_{2r-2}}\in \DP$
for all $i,j=1,\dots ,r$. \N

In the last two results, Proposition \ref{prop:square} and Corollary
\ref{cor:squares}, the special case with
$\left(T{}_i\!\times_i\! T\right)_{a_1\dots a_{2r-2}}=0$, which is in $\DP$,
has been excluded. Similar results apply for this extreme case, but they
need some refinement.
\begin{prop}
$\left(T{}_i\!\times_i\! T\right)_{a_1\dots a_{2r-2}}=0$ for some
$i=1,\dots ,r \, \Longleftrightarrow$ there is a {\em null} vector $\vec{k}$
such that
$T_{a_{1}...a_{r}}=k_{a_i}t_{a_1\dots a_{i-1}a_{i+1}\dots a_r}$ for some
tensor $t_{a_1\dots a_{r-1}}$.
\label{prop:zerosquare}
\end{prop}
\P Using the same notation as in Proposition \ref{prop:square} we have that
all the $v_b$ are causal and furthermore, as $v_b\tilde{v}^b=0$, all of them
are orthogonal to each other. This means that all of them must be proportional
to a null vector $v_b \propto k_b$, and the result follows. The converse is
immediate.\N

Therefore, it can happen that
$\left(T{}_i\!\times_i\! T\right)_{a_1\dots a_{2r-2}}=0$ and therefore is in
$\DP$ and yet neither $T_{a_{1}...a_{r}}$ nor $-T_{a_{1}...a_{r}}$ is in
$\DP$.
It is enough that $t_{a_1\dots a_{r-1}}\notin \DP\cup -\DP$. Nevertheless,
we have the following result.
\begin{coro}
$\left(T{}_i\!\times_i\! T\right)_{a_1\dots a_{2r-2}}=0$ for {\em all}
$i=1,\dots ,r \Longleftrightarrow T_{a_{1}...a_{r}}=k_{a_1}\dots n_{a_r}$
for a set of $r$ null vectors $\vec{k},\dots ,\vec{n}$.

Therefore, $\left(T{}_i\!\times_i\! T\right)_{a_1\dots a_{2r-2}}=0$
for {\em all} $i=1,\dots ,r \Longrightarrow  T_{a_{1}...a_{r}}
\in \DP\cup -\DP$.
\label{cor:zerosquare}
\end{coro}
\P The first part follows from repeated application of
Proposition \ref{prop:zerosquare}. Then, depending on how many of the
null vectors $\vec{k},\dots ,\vec{n}$ are future-pointing, either
$T_{a_{1}...a_{r}}$ or $-T_{a_{1}...a_{r}}$ is in $\DP$.\N
\begin{coro}
Assume that $T_{a_{1}...a_{r}}$ is completely symmetric.
Then, $\left(T{}_i\!\times_j\! T\right)_{a_1\dots a_{2r-2}} =0
\Longleftrightarrow T_{a_{1}...a_{r}}=f k_{a_1}\dots k_{a_r}$ for some
future-pointing null vector $\vec{k}$.
\label{cor:zerosquaresym}
\end{coro}
\P If $T_{a_{1}...a_{r}}$ is completely symmetric then all the products
$\left(T{}_i\!\times_j\! T\right)_{a_1\dots a_{2r-2}}$ are the same. Thus,
from
Corollary \ref{cor:zerosquare} and the symmetry of $T_{a_{1}...a_{r}}$ the
result follows.\N

Of course, under the assumptions of Corollary \ref{cor:zerosquaresym},
$T_{a_{1}...a_{r}} \in \DP\cup -\DP$.
On the other hand, we have
\begin{coro}
If $T_{a_{1}...a_{r}}$ is completely antisymmetric and
$\left(T{}_i\!\times_i\! T\right)_{a_1\dots a_{2r-2}} = 0$ then
$T_{a_{1}...a_{r}}=0$.
\label{cor:zerosquareasym}
\end{coro}
\P If $T_{a_{1}...a_{r}}$ is completely antisymmetric again all the products
$\left(T{}_i\!\times_j\! T\right)_{a_1\dots a_{2r-2}}$ are the same. Thus,
from
Corollary \ref{cor:zerosquare} and the antisymmetry of $T_{a_{1}...a_{r}}$ the
only possibility is $T_{a_{1}...a_{r}}=0$.\N

\section{Superenergy tensors}
In the previous section we have defined the set $\DP$ and analyzed its
general properties. However, we must still face the question of how
general is the class $\DP$ and how we can build such causal tensors.
Actually this has been already solved and the result is that, given an
{\em arbitrary} tensor $A_{c_1 \dots c_m}$, there is a general procedure
to construct its ``positive square'': a tensor quadratic in $A_{c_1 \dots
c_m}$
and with the dominant property. This general procedure was introduced in
\cite{S1} and extensively considered in \cite{S2}, and the positive tensors
thus constructed receive the generic name of ``super-energy tensors'' (due
to historical reasons \cite{S2}). In what follows, we recall here
the definition of a general superenergy tensor (see section 3 of \cite{S2}).

Let $A_{c_1 \dots c_m}$ be an {\em arbitrary} rank-$m$ tensor.
Let $[n_1]$ denote the set of indices containing $c_1$ and all other
indices $c_j$ such that $A_{c_1 \dots c_m}$ is anti-symmetric in
$c_1 c_j$. The number $n_1$ is the number of indices in $[n_1]$.
Then $[n_2]$ is the next set formed from anti-symmetries with $c_2$
(or $c_3$ if $c_2$ is already in $[n_1]$ and so on). Note that
$1\le n_i\le N$ for each $i$. In this way
$c_1, \dots ,c_m$ are divided into $r$ blocks $[n_1], \dots ,[n_r]$
with $n_1+ \dots +n_r =m$. We can therefore consider
$A_{c_1 \dots c_m}$ as an $r$-fold $(n_1, \dots ,n_r )$-form and
we write $A_{c_1 \dots c_m}=A_{[n_1] \dots [n_r]}$.

There are $2^r$ different (multiple) Hodge duals of
$A_{[n_1] \dots [n_r]}$. The dual with respect to the block $[n_1]$
is denoted $A_{\stackrel{*}{[N-n_1]}[n_2] \dots [n_r]}$, with respect to
$[n_2]$ by $A_{[n_1]\stackrel{*}{[N-n_2]} \dots [n_r]}$, the dual with
respect to $[n_1]$ and $[n_2]$ by
$A_{\stackrel{*}{[N-n_1]}\stackrel{*}{[N-n_1]} \dots [n_r]}$ and so on.
Note that different duals may be tensors of different rank but all
duals are $r$-folded forms.
Denote by $(A_{ \cal P })_{[\ ]\dots [\ ]}$ , ${\cal P} =
1,2,\dots ,2^r$,
all the possible duals, where ${\cal P} \equiv 1+s_1+2s_2+\dots
+2^{r-1}s_r$ with $s_i=0$ if there is no dual with respect to
the block $[n_i]$ and $s_i =1$ if there is a dual with respect to this
block (so $A_{1}=A$ and $A_{2^r}$ is the $r$-fold form where the
dual has been taken with respect to all blocks).

In order to define the superenergy tensor of $A_{c_1 \dots c_m}$
we need a product $\odot$ of an $r$-fold form by itself
resulting in a $2r$-tensor. Let $\tilde A_{c_1 \dots c_m}$
be the tensor obtained by permuting the indices in
$A_{c_1 \dots c_m}$ so that the $n_1$ first indices in
$\tilde A_{c_1 \dots c_m}$ are precisely the indices in the block $[n_1]$,
the following $n_2$ indices are the ones in $[n_2]$ and so on.
Now define the product $\odot$ by\footnote{The signs in this
formula arise here because of the different choice of signature with
respect to \cite{S2}.}
\be
(A\odot A )_{a_1 b_1 \dots a_r b_r}=
\left(\prod_{\U=1}^{r}\frac{(-1)^{n_\U-1}}{(n_\U-1)!}\right)\,
\tilde{A}_{a_1c_2\dots c_{n_1},\dots ,a_rd_2\dots d_{n_r}}
\tilde{A}_{b_1}{}^{c_2\dots c_{n_1},}{}_{\dots ,b_r}
{}^{d_2\dots d_{n_r}} \, . \label{semi-sq}
\ee
 From each block in $A_{[n_1] \dots [n_r]}$ two indices are obtained
in $(A\odot A )_{a_1 b_1 \dots a_r b_r}$. We can form
$(A_{\cal P}\odot A_{\cal P} )_{a_1 b_1 \dots a_r b_r}$ for any
$\cal P$ but observe that $A_{[n_1] \dots [n_r]}$ could contain $N$-blocks
(with dual 0-blocks) for which the expression (\ref{semi-sq}) has no meaning.
Therefore, assuming $1\le n_i\le N-1$ for all $i$,
we make the following definition.
\begin{defi}
The \underline{superenergy tensor} of
$A_{c_1 \dots c_m}$ is defined to be
\[
T_{a_1 b_1 \dots a_r b_r} \{ A\} ={1\over 2}\sum_{{\cal P}=1}^{2^r}
(A_{\cal P}\odot A_{\cal P} )_{a_1 b_1 \dots a_r b_r}
\]
\label{def:se}
\end{defi}
Observe that any dual $A_{\cal P}$ of the original tensor $A=A_1$
generates the same superenergy tensor. We note that
\[
T_{a_1 b_1 \dots a_r b_r} \{ A\}=
T_{(a_1 b_1 )\dots (a_r b_r)} \{ A\}
\]
and if $A_{[n_1] \dots [n_r]}$ is symmetric with respect to two blocks
$[n_{\U}]$ and $[n_{\tilde{\U}}]$, then $T_{a_1 b_1 \dots a_r b_r}$ is symmetric
with respect to the pairs $a_{\U} b_{\U}$ and $a_{\tilde{\U}} b_{\tilde{\U}}$.
Important is also the
property that $T_{a_1 b_1 \dots a_r b_r} \{ A\}=0$ if and only if
$A_{c_1 \dots c_m}=0$ \cite{S2}.

Another property of superenergy tensors is:
\begin{pr}
If $A_{[n_1] \dots [n_r]}=B_{[n_1] \dots [n_s]}C_{[n_{s+1}] \dots [n_r]}$
for some $s$- and $(r-s)$-folded forms $B_{[n_1] \dots [n_s]}$
and $C_{[n_{s+1}] \dots [n_r]}$, then
$$
T_{a_1 b_1 \dots a_r b_r}
\left\{B_{[n_1] \dots [n_s]}C_{[n_{s+1}] \dots [n_r]} \right\}=
T_{a_1 b_1 \dots a_s b_s} \left\{B_{[n_1] \dots [n_s]} \right\}
T_{a_{s+1} b_{s+1} \dots a_r b_r} \left\{C_{[n_{s+1}] \dots [n_r]} \right\}.
$$
\label{pr:seprod}
\end{pr}
\P This follows at once from Definition \ref{def:se} because the $2^s$
terms and the $2^{r-s}$ terms of
$T_{a_1 b_1 \dots a_s b_s} \left\{B_{[n_1] \dots [n_s]} \right\}$ and
$T_{a_{s+1} b_{s+1} \dots a_r b_r} \left\{C_{[n_{s+1}] \dots [n_r]} \right\}$,
respectively, produce precisely the $2^r$ terms needed.\N
\begin{defi}
An r-fold form $A_{[n_1] \dots [n_r]}$ is said to be \underline{decomposable}
if there are $r$ forms $\O^{(\U)}_{a_1\dots a_{n_{\U}}}=
\O^{(\U)}_{[a_1\dots a_{n_{\U}}]}$ ($\U=1\dots r$)
such that $\tilde{A}_{[n_1] \dots [n_r]}=\left(\O^{(1)}\otimes \dots \otimes
\O^{(r)}\right)_{[n_1] \dots [n_r]}$.
\label{def:decom}
\end{defi}
\begin{coro}
If $A_{[n_1] \dots [n_r]}$ is decomposable, then
$$
T_{a_1 b_1 \dots a_r b_r}\left\{A_{[n_1] \dots [n_r]} \right\}=
T_{a_1b_1}\left\{\O^{(1)}_{[n_1]} \right\}\dots
T_{a_rb_r}\left\{\O^{(r)}_{[n_r]} \right\}.$$
\label{cor:decom}
\end{coro}
\P This is evident from Property \ref{pr:seprod} and
Definition \ref{def:decom}.\N

The last result shows that rank-2 superenergy tensors may be used as basic
set to build up more general superenergy tensors in many occasions.
Actually, we will later be interested (for other important
reasons) in rank-2 tensors, specially those in $\DP$.

It is remarkable that, after expanding all duals in the Definition
\ref{def:se},
one obtains an explicit expression for the general superenergy tensor which is
{\it independent} of the dimension $N$, see \cite{S2}. In the case of a
general $p$-form $\Omega _{a_{1}...a_{p}}$, the rank-2 superenergy tensor
becomes \cite{S2}
\be
T_{ab}\{\Omega_{[p]} \}=\frac{(-1)^{p-1}}{(p-1)!}\left[
\Omega_{aa_{2}...a_{p}}\Omega_{b}{}^{a_{2}...a_{p}}-
\frac{1}{2p}(\Omega \cdot \Omega) g_{ab}\right]\, .\label{sep}
\ee
Here we have used a notation which will be useful on many occasions: for
any two tensors of the same rank $a_{a_1\dots a_j}$ and $b_{a_1\dots a_j}$,
we write $a_{a_1\dots a_j}b^{a_1\dots a_j}\equiv a\cdot b$,
i.e. we have contracted over all indices in order.

In the Definition \ref{def:se} we assumed that there were no $N$-blocks.
The expression (\ref{sep}) however is perfectly well defined for an
$N$-form. If $\Omega _{a_{1}...a_{N}}=f\eta_{a_{1}...a_{N}}$ where
$\bm\eta$ is the canonical volume form and $f$ a scalar, then (\ref{sep}) gives
\be
T_{ab}\{\Omega_{[N]} \}={1\over 2}f^2 g_{ab}\, .\label{seN}
\ee

If we combine (\ref{seN}) with Property \ref{pr:seprod} the Definition
\ref{def:se} is naturally extended to include $N$-blocks:

\begin{defi}
The \underline{superenergy tensor} of
$A_{c_1 \dots c_m}=A_{[n_1] \dots [n_{r-1}][N]}=
B_{[n_1] \dots [n_{r-1}]}\Omega_{[N]}$, with
$\Omega _{a_{1}...a_{N}}=f\eta_{a_{1}...a_{N}}$ is defined to be
\[
T_{a_1 b_1 \dots a_{r-1} b_{r-1} a_r b_r} \{ A\} ={1\over 2}f^2
T_{a_1 b_1 \dots a_{r-1} b_{r-1}} \{ B\}g_{a_r b_r}
\]
\label{def:seN}
\end{defi}

This definition is to be understood recursively, if there are $q$
$[N]$-blocks one continues until a tensor
$T_{a_1 b_1 \dots a_{r-q} b_{r-q}} \{ B\}$ given by Definition \ref{def:se}
is obtained.

We note that the tensor obtained by taking the dual of $A_{[n_1] \dots [n_{r-1}][N]}$
with respect to the $N$-block to get a 0-block does not have the same superenergy tensor
as $A_{[n_1] \dots [n_{r-1}][N]}$ has, the difference being the
$g_{a_r b_r}$. This is a special situation only for $N$-blocks
and it is the price one has to pay to extend the definition.
The advantages however will be seen in a more consistent presentation
of several definitions and results, the first being the following
definition.

\begin{defi}
The set $\SE$ is the set of all superenergy tensors according to
Definitions \ref{def:se} and \ref{def:seN}. By $-\SE$ we denote the set of tensors
such that $-T_{a_{1}...a_{r}}\in \SE$. The sets $\SE_n$ and $-\SE_n$
will denote the classes of rank-n tensors in $\SE$ and $-\SE$,
respectively.
\label{def:SE}
\end{defi}

The metric is an essential element in this set.
In \cite{B2} it was shown that the metric $g_{ab}$ is not a superenergy tensor
of any $p$-form $\Omega_{[p]}$ with $1\le p\le N-1$ so without the extended
definition elements of the form $f^2g_{ab}$ would have to be added artificially
to $\SE$.

A fundamental result is that superenergy tensors always have the
dominant property.
\begin{theo}
$\SE\subset \DP$.
\label{th:dp}
\end{theo}
\P The first proof for 4 dimensions was given in \cite{B2} and used
spinors. In
arbitrary dimension the first proof is in \cite{S2} while a proof that uses
Clifford algebras and which is also valid in arbitrary dimension was
presented in \cite{PP}. Of course, $-\SE\subset -\DP$. \N

It is important to remark that the superenergy tensor
$T_{a_1 b_1 \dots a_r b_r} \{ A\}$ and its derived tensors by permutation of
indices are the {\it{only}} (up to linear combinations) tensors quadratic in
$A_{c_1 \dots c_m}$ and with the dominant property \cite{S2}.
Therefore, there is a unique (up to a
proportionality factor) {\it completely symmetric} tensor in $\DP$ which is
quadratic in $A_{c_1 \dots c_m}$, and this is simply
$T_{(a_1 b_1 \dots a_r b_r)} \{ A\}$ \cite{S2}.

In $N=4$, the superenergy tensor of a 2-form $F_{ab}=F_{[ab]}$ is its
Maxwell energy-momentum tensor, and the superenergy tensor of an exact
1-form $d\phi$ has the form of the energy-momentum tensor for a massless
scalar field $\phi$. If we compute the superenergy tensor of the Riemann
tensor, which is a double symmetrical (2,2)-form, we get the so-called
Bel tensor \cite{Bel}. The superenergy tensor of the Weyl curvature tensor
is the well-known Bel-Robinson tensor \cite{B,Bel2}. For these and other
interesting physical examples see \cite{B2,S2,S3}. The dominant property of
the Bel-Robinson tensor was used by Christodoulou and Klainerman \cite{CK}
in their study of the global stability of Minkowski spacetime, and in
\cite{BoS} to study the causal propagation of gravity in vacuum. The dominant
property of more general superenergy tensors was used in \cite{BS}
to find criteria for the causal propagation of fields on
Lorentzian spacetimes of $N$ dimensions.

In order to study relations between $\SE_2$ and $\DP$, and to see how
$\SE_2$ builds up $\DP$, we prove now
some results for rank-2 tensors. First, we need a very simple Lemma to fix the notation.
\begin{lem}
For any $T_{ab}$ we have that $\left(T{}_1\!\times_1\! T\right)_{ab}$ and
$\left(T{}_2\!\times_2\! T\right)_{ab}$ are symmetric. Furthermore, if
$s_{ab}=s_{(ab)}$ and $t_{ab}=t_{(ab)}$ are symmetric, then
$$
\left(s\, {}_1\!\!\times_1 t\right)_{ab}=
\left(s\, {}_1\!\!\times_2 t\right)_{ab}=
\left(s\, {}_2\!\!\times_1 t\right)_{ab}=
\left(s\, {}_2\!\!\times_2 t\right)_{ab}
\equiv \left(s \times t\right)_{ab}\, ;
$$
if $F_{ab}=F_{[ab]}$ and $G_{ab}=G_{[ab]}$ are antisymmetric, then
$$
\left(F{}_1\!\times_1\! G\right)_{ab}=-\left(F{}_1\!\times_2\! G\right)_{ab}=
-\left(F{}_2\!\times_1\! G\right)_{ab}=\left(F{}_2\!\times_2\! G\right)_{ab}
\equiv \left(F \times G\right)_{ab}\, ;
$$
finally in the mixed case
$$
\left(s\, {}_1\!\!\times_1 F\right)_{ab}=
-\left(s\, {}_1\!\!\times_2 F\right)_{ab}=
\left(s\, {}_2\!\!\times_1 F\right)_{ab}=
-\left(s\, {}_2\!\!\times_2 F\right)_{ab}
\equiv \left(s \times F\right)_{ab}
$$
so that in the three cases the simple $\times$-notation will be used.
\label{lem:x}
\end{lem}
\P First, $\left(T{}_1\!\times_1\! T\right)_{ab}=T_{ca}T^{c}{}_b$ which is
obviously symmetric in $ab$, and analogously for ${}_2\!\times_2$. If
$s_{ab},r_{ab}$ are symmetric then
$s_{ca}r^{c}{}_b=s_{ac}r^{c}{}_b=s_{ac}r_b{}^c$ and
similarly for the other cases.\N
\begin{lem}
If $F_{ab}=F_{[ab]}\neq 0$ is a 2-form, then
$\left(F \times F\right)_{ab}\notin \DP$.
\label{lem:F}
\end{lem}
\P If $\left(F \times F\right)_{ab}$ were in $\DP$ and non-zero, then from
Proposition \ref{prop:square} $F_{ab}$ should be in $\DP \cup -\DP$, which
is impossible due to Lemma \ref{lem:skew}. If $\left(F \times
F\right)_{ab}=0$,
then from Corollary \ref{cor:zerosquareasym} it follows
that $F_{ab}=0$. \N

Notice that, still, $\left(F \times F\right)_{ab}$ can certainly be in $-\DP$.

\begin{prop}
In dimension $N\leq 4$, $T_{ab}\in \SE_2 \Longrightarrow
(T\times T)_{ab}=h^{2}g_{ab}$. On the other hand, if $N>4$ there exist
tensors $T_{ab}\in \SE_2$ such that $(T\times T)_{ab}$
is not proportional to the metric.
\label{prop:TxT}
\end{prop}
\P If $T_{ab}=hg_{ab}$ then $(T\times T)_{ab}=h^{2}g_{ab} $ so the property is
trivial. As the superenergy of a
$p$-form is the same as the superenergy of its dual $(N-p)$-form, we just have
to confirm the proposition for 1-forms for $N\leq 3$, and for 1-forms and
2-forms for $N=4$. By (\ref{sep}), the superenergy tensor of a 1-form $J_{a}$
in any dimension $N$ is
\be
T_{ab}\{J_{[1]}\}=J_{a}J_{b}-\frac{(J\cdot J)}{2}g_{ab} \label{se1}
\ee
and this gives
\b*
\left(T\{J_{[1]}\} \times T\{J_{[1]}\}\right)_{ab}=
\frac{(J\cdot J)^{2}}{4}g_{ab}
\e*
so $\left(T\{J_{[1]}\} \times T\{J_{[1]}\}\right)_{ab}$
is proportional to the metric in any dimension. For a
2-form $F_{ab}$, the superenergy tensor reads
\be
T_{ab}\{F_{[2]}\}=-F_{ac}F_{b}{}^{c}+(1/4)(F\cdot F) g_{ab} \label{se2}
\ee
which again, by (\ref{sep}), holds in any $N$. Now, if $N=4$, and {\em only} in this case,
a very well-known result is (see, e.g., \cite{L1,L2,MW,PR,R})
\be
\left(T\{F_{[2]}\} \times T\{F_{[2]}\}\right)_{ab}=\frac{1}{16}
\left[(F\cdot F)^2 + (F\cdot *F)^2\right]\, g_{ab} \hspace{1cm}
\mbox{if $N=4$} \label{rai}
\ee
where $*F_{ab}$ is the 2-form dual to $F_{ab}$ in 4 dimensions.
This is the basis of the Rainich theory \cite{R,MW,PR} and, as
was pointed out by Lovelock \cite{L1,L2}, formula (\ref{rai}) is an explicit
example of a {\em dimensionally-dependent} identity, being valid only in
$N=4$.

Not even by changing the proportionality factor on the righthand side
the above expression (\ref{rai}) holds in $N>4$. To check it,
we can construct explicit counterexamples. Let
$\left\{\bm{e}_{0},\bm{e}_{1},...,\bm{e}_{N-1}\right\}$ be an orthonormal
basis and let $F_{ab}=(\bm{e}_0\wedge \bm{e}_1)_{ab}+
(\bm{e}_2\wedge \bm{e}_3)_{ab}$.
Then the computation of (\ref{se2}) gives
$T_{ab}\{F_{[2]}\}=(\bm{e}_0\otimes \bm{e}_0)_{ab}-
(\bm{e}_1\otimes \bm{e}_1)_{ab}+
(\bm{e}_2\otimes \bm{e}_2)_{ab}+(\bm{e}_3\otimes \bm{e}_3)_{ab}$ from
where one immediately obtains
$\left(T\{F_{[2]}\} \times T\{F_{[2]}\}\right)_{ab}=
(\bm{e}_0\otimes \bm{e}_0)_{ab}-
(\bm{e}_1\otimes \bm{e}_1)_{ab}-
(\bm{e}_2\otimes \bm{e}_2)_{ab}-
(\bm{e}_3\otimes \bm{e}_3)_{ab}$
which is (proportional to) the metric in 4 but not higher dimension.\N

Thus, for $N\leq 4$, $\SE_2$ is the set of tensors with the property
that $(T\times T)_{ab}$ is porportional to the metric, but this is not true
for $N>4$. The natural question arises of which super-energy tensors
satisfy this property in arbitrary $N$. This is going to be answered now, and
in a more complete way in the next section. The generalization of the
algebraic Rainich condition (\ref{rai})
will be dealt with in the last section.

Recall that a $p$-form $\Omega_{a_{1}...a_{p}}=\Omega_{[a_{1}...a_{p}]}$
is called \underline{simple} \cite{Sc,PR}
if it is a product of $p$ linearly independent
1-forms $\bm{\omega}^{1},\dots ,\bm{\omega}^{p}$, i.e.
$\Omega _{a_{1}...a_{p}}=
(\bm{\omega}^{1}\wedge ...\wedge \bm{\omega}^{p})_{a_{1}...a_{p}}$.
By standard techniques, the set $\bm{\omega}^{1},\dots ,\bm{\omega}^{p}$
can be chosen to be orthogonal by simply taking the appropriate linear
combinations $\bm{\omega}^{i'}=a^{i'}_{j}\bm{\omega}^{j}$ with
$\det(a^{i'}_{j})=1$, because
$(\bm{\omega}^{1}\wedge ...\wedge \bm{\omega}^{p})_{a_{1}...a_{p}}=
(\bm{\omega}^{1'}\wedge ...\wedge \bm{\omega}^{p'})_{a_{1}...a_{p}}$.
A $p$-form $\Omega_{[p]}$ is simple if and only if
$\Omega_{\stackrel{*}{[N-p]}}$ is simple, and if and only if
$\left(\Omega_{[p]}\, {}_1\!\!\times_1
\Omega_{\stackrel{*}{[N-p]}}\right)_{a_1\dots a_{N-2}}=0$,
see e.g. \cite{Sc,PR}.
\begin{defi}
We denote by $\SS$ the set of superenergy tensors of {\em simple}
$p$-forms. Observe that
$\SS\subset \SE_2$. We define $-\SS$ as usual.
\label{def:SS}
\end{defi}
\begin{prop}
$T_{ab}\in \SS \Longrightarrow (T\times T)_{ab}=h^{2}g_{ab}$.
\label{prop:txt}
\end{prop}
\P From the proof of Proposition \ref{prop:TxT} the result is already
proved for $fg_{ab}$ and for the superenergy tensor
$T_{ab}\{\Omega_{[1]}\}$ of any 1-form. Using (\ref{sep}) for the
superenergy tensor of a general $p$-form $\Omega _{a_{1}...a_{p}}$,
a straightforward calculation gives
\b*
[(p-1)!]^2\left(T\{\Omega_{[p]}\}\times T\{\Omega_{[p]} \}\right)_{ab}=
\hspace{3cm}\\
=\Omega_{a}{}^{a_{2}...a_{p}}\Omega _{b b_{2}...b_{p}}
\Omega_{c a_{2}...a_{p}}\Omega^{c b_{2}...b_{p}}-
\frac{\Omega \cdot \Omega }{p}\Omega_{aa_{2}...a_{p}}
\Omega_{b}{}^{a_{2}...a_{p}}+
\frac{(\Omega \cdot \Omega )^{2}}{4p^{2}}g_{ab}
\e*
Now, in the first term
$\Omega _{b b_{2}...b_{p}}
\Omega_{c a_{2}...a_{p}}\Omega^{c b_{2}...b_{p}}=
\Omega _{b[b_{2}...b_{p}}\Omega_{c]a_{2}...a_{p}}\Omega^{c b_{2}...b_{p}}$.
By \cite{Sc} (p.23) (or \cite{PR} (p.165)) we have
$\Omega _{b[b_{2}...b_{p}}\Omega_{c]a_{2}...a_{p}}=
(1/p)\Omega_{ba_{2}...a_{p}}\Omega _{cb_{2}...b_{p}}$
{\it{if and only if}} $\Omega_{a_{1}...a_{p}} $ {\it{is simple}}.
Thus, in this case we are left with
\be
\left(T\{\Omega_{[p]}\}\times T\{\Omega_{[p]} \}\right)_{ab}=
\frac{(\Omega \cdot \Omega )^{2}}{(2\, p!)^{2}}\, g_{ab} \label{ssxss}
\ee
which proves the proposition.\N
\begin{prop}
If $\Omega _{a_{1}...a_{p}}=
(\bm{\omega}^{1}\wedge ...\wedge \bm{\omega}^{p})_{a_{1}...a_{p}}$ is a
simple $p$-form, then $\vec{\omega}^1,\dots ,\vec{\omega}^p$ are eigenvectors,
all with the same eigenvalue $\left[(-1)^{p-1}(\Omega \cdot \Omega)\right]/(2p!)$,
of its superenergy tensor $T_{ab}\{\Omega_{[p]} \}\in \SS$.
\label{prop:eigen}
\end{prop}
\P The case $p=N$ is trivial so assume $p<N$.
The dual $\Omega_{\stackrel{*}{[N-p]}}$ of $\Omega_{[p]}$ is obviously
orthogonal to any of the $\vec{\omega}^1,\dots ,\vec{\omega}^p$. But the
superenergy tensors $T_{ab}\{\Omega_{[p]} \}$ and
$T_{ab}\{\Omega_{\stackrel{*}{[N-p]}} \}$ are identical, so that using
the explicit expression (\ref{sep}) for
$T_{ab}\{\Omega_{\stackrel{*}{[N-p]}} \}$ and contracting with
$\vec{\omega}^i$ for $i=1,\dots ,p$ we get
$$
\omega^i_a T_b{}^a\{\Omega_{\stackrel{*}{[N-p]}} \}=
\omega^i_a T_b{}^a\{\Omega_{[p]} \}=
\frac{(-1)^{p-1}}{2\, p!}(\Omega \cdot \Omega)
\omega^i_b
$$
for any $i$.\N

Recall finally that a $p$-form $\Omega_{[p]}$ is called \underline{null}
if it is simple and $(\Omega \cdot \Omega )=0$. Then, $\Omega_{[p]}$
defines canonically a null direction $\vec{k}$ such that
$(\bm{k}\wedge \Omega_{[p]})_{a_1\dots a_{p+1}}=0$ and
$\left(\bm{k}\wedge
\Omega_{\stackrel{*}{[N-p]}}\right)_{a_1\dots a_{N-p+1}}=0$. Equivalently,
$\Omega_{[p]}$ can be written in the form
$\Omega_{[p]}=(\bm{k}\wedge
\bm{\omega}^{2}\wedge ...\wedge \bm{\omega}^{p})_{a_{1}...a_{p}}$ where the
$(p-1)$ 1-forms $\bm{\omega}^{2},\dots ,\bm{\omega}^{p}$ are mutually
orthogonal and orthogonal to $\vec{k}$.
\begin{defi}
We denote by $\NS$ the set of superenergy tensors of {\em null}
$p$-forms. Obviously, $\NS\subset \SS\subset \SE_2 \subset \DP$. $-\NS$ is
defined as usual.
\label{def:NS}
\end{defi}
\begin{coro}
$T_{ab}\in \NS \Longrightarrow (T\times T)_{ab}=0$ and $T_{ab}=fk_a k_b$ where
$\vec{k}$ is null and $f>0$.
\label{cor:NxN}
\end{coro}
\P As $T_{ab}=T_{ab}\{\Omega _{[p]}\}$ for a null $p$-form $\Omega_{[p]}$, so
that $(\Omega\cdot \Omega)=0$, from (\ref{ssxss}) we get
$(T\times T)_{ab}=0$. Furthermore, as $\Omega_{a_{1}\dots a_{p}}=(\bm{k}\wedge
\bm{\omega}^{2}\wedge ...\wedge \bm{\omega}^{p})_{a_{1}...a_{p}}$ where
$k_a$ is its canonical null direction, a simple calculation produces
$T_{ab}=f k_a k_b$ where $f\equiv
(-1)^{p-1}(\bm{\omega}^{2}\cdot \bm{\omega}^{2})\cdots
(\bm{\omega}^{p}\cdot \bm{\omega}^{p})$ which is positive.\N
\begin{coro}
$T_{ab}\in \SS\setminus\NS \Longrightarrow T_{ab}$ has N independent
eigenvectors, p of them with eigenvalue $\left[(-1)^{p-1}(\Omega \cdot \Omega)
\right]/(2p!)$ and (N-p) of them with the opposite eigenvalue.
$T_{ab}\in \NS \Longrightarrow T_{ab}$ has N-1 independent
eigenvectors, all with zero eigenvalues. It has
a unique null eigenvector.
\label{cor:Nevec}
\end{coro}
\P Again $p=N$ is trivial so assume $p<N$.
As $\Omega_{\stackrel{*}{[N-p]}}$ is simple if $\Omega _{[p]}$
is simple, and as $T_{ab}\{\Omega_{[p]} \}=
T_{ab}\{\Omega_{\stackrel{*}{[N-p]}} \}$, the $(N-p)$ 1-forms that generate
$\Omega_{\stackrel{*}{[N-p]}}$ are also eigenvectors of
$T_{ab}\{\Omega_{[p]} \}$. From (\ref{sep}) one immediately finds
the eigenvalues $(-1)^{p}(\Omega \cdot \Omega)/(2\, p!)$.
The case $T_{ab}\in \NS$ is trivial from Corollary \ref{cor:NxN}.
\N

\section{Maps preserving the null cone and $\DP$}
In this section we are going to show two important properties of the set
$\SS$:
on one hand, its elements are the basic building blocks of all rank-2 tensors
in $\DP$, and on the other they define maps which leave the null cone
invariant.
The converse of this result also holds but is left for the last section.
\begin{defi}
We say that $T_{a}{}^b$ defines a \underline{null-cone preserving map}
if $k^a T_{a}{}^b$ is null for any null vector $\vec{k}$. A map that
preserves the null cone is said to be \underline{orthochronus} (respectively
\underline{time reversal}) if it keeps (resp. reverses) the cone's
time orientation, and is called \underline{proper}, \underline{improper} or
\underline{singular} if $\det(T_{a}{}^b)$ is positive, negative, or zero,
respectively. If the map is proper and orthochronus then it is called
\underline{restricted}. A null-cone preserving map is \underline{involutory}
if $T_{a}{}^b=(T^{-1}){}_{a}{}^b$, and \underline{bi-preserving} if
$T_{a}{}^b k_b$ is also null for any null 1-form $\bm{k}$.
\label{def:maps}
\end{defi}
Most of the above terminology is taken from that of Lorentz
transformations, see e.g. \cite{PR}. Notice that involutory null-cone
preserving maps are necessarily non-singular. In order to characterize
all these maps and relate them to $\SS$ we first recall a simple result.
\begin{lem}
$T_{(ab)}=fg_{ab} \Longleftrightarrow T_{ab}k^{a}k^{b}=0$
for any $k^{a}$ that is null.
\label{lem:fg}
\end{lem}
\P The implication from left to right is trivial. Conversely,
if $T_{ab}k^{a}k^{b}=T_{(ab)}k^{a}k^{b}=0$, take an orthonormal basis
$\left\{\vec{e}_{0},\vec{e}_{1},...,\vec{e}_{N-1}\right\}$ with
a timelike $\vec{e}_{0}$. Using first as null $\vec{k}$ the vectors
$\vec{e}_{0}\pm \vec{e}_{i}$ for $i=1\dots ,N-1$ one immediately deduces
$T_{(0i)}=0$ and $T_{00}+T_{ii}=0$ for each $i$. Using then as null $\vec{k}$
the vectors
$\vec{e}_{0}\pm \cos\alpha \, \vec{e}_{i}\pm \sin\alpha\, \vec{e}_{j}$
for $i,j=1\dots ,N-1$ one gets $T_{(ij)}=0$ for all $i\neq j$. \N

The following Lemma gives important geometrical interpretations
to some results.
\begin{lem}
$(T {}_2\!\times_2 T)_{ab}=fg_{ab} \Longleftrightarrow T_{a}{}^b$ is a
null cone preserving map.
\label{lem:TxT=fg}
\end{lem}
\P The basic formula is $(T {}_2\!\times_2 T)_{ab}k^{a}k^{b}=
(T_{ac}k^{a})(T_{b}{}^{c}k^{b})$. If
$(T{}_2\!\times_2 T)_{ab}=fg_{ab}$ then for any null $k^b$ we have
that $T_{ab}k^{a}$ must be null. Conversely, if $T_{ab}k^{a}$ is null for any
$k^{a}$ that is null, and given that $(T {}_2\!\times_2 T)_{ab}$ is symmetric
according to Lemma \ref{lem:x}, then by Lemma \ref{lem:fg}
$(T{}_2\!\times_2 T)_{ab}$ must be proportional to the metric. \N
\begin{coro}
If $(T {}_2\!\times_2 T)_{ab}=fg_{ab}$ then $f\ge 0$ (for $N\neq 2$).
\label{cor:f>0}
\end{coro}
\P From Lemma \ref{lem:TxT=fg} we know that $T_{ab}k^{a}$ is null for any
$k^{a}$ that is null. If $f$ were negative, then for any null and
future-pointing vectors $k^{a}$ and $n^b$ we would have
$(T_{ac}k^{a})(T_{b}{}^{c}n^{b})=f\, k_{c}n^{c}<0$, so that {\em any}
two null vectors of type $T_{ab}k^{a}$ and $T_{ab}n^{a}$ would have opposite
time orientations. But this is evidently impossible for {\em all} the null
vectors of type $T_{ab}k^{a}$ {\it unless} there are only two, that
is, $N=2$.\N

Similar results can be shown for the product $(T{}_1\!\times_1 T)_{ab}$.
However, they are mainly redundant because of the following
\begin{lem}
In $N>2$, $(T{}_2\!\times_2 T)_{ab}=f g_{ab}\neq 0 \Longleftrightarrow f>0$ and
$(T{}_1\!\times_1 T)_{ab}=f g_{ab} \Longleftrightarrow T_a{}^b$ defines
a non-singular null-cone preserving map. A fortiori, all
non-singular maps preserving the null cone are automatically bi-preserving,
proportional to an $N$-dimensional Lorentz transformation, and in
$\DP\cup -\DP$.
\label{lem:22=11}
\end{lem}
\P If $(T{}_2\!\times_2 T)_{ab}=f g_{ab}\neq 0$,
from Corollary \ref{cor:f>0} we get
$f>0$ so that we can define $L_{ab}\equiv \frac{1}{\sqrt{f}}T_{ab}$
and the condition becomes $g_{cd}L_{a}{}^c L_{b}{}^d=g_{ab}$.
This means that $L_{a}{}^b$ is a Lorentz
transformation (ergo non-singular), which as is well-known also satisfies
$g_{cd}L^c{}_{a}L^d{}_{b}=g_{ab}$, see e.g. \cite{Sc,PR}. This is exactly
$(T{}_1\!\times_1 T)_{ab}=f g_{ab}$. Now, a reasoning identical to that in
the proof of Lemma \ref{lem:TxT=fg} implies that $T_{ab}k^{b}$ is null for any
$k^{a}$ that is null, that is, $T_{ab}$ is bi-preserving. Finally, as $f>0$,
$(T{}_2\!\times_2 T)_{ab}\in \DP$ so that from Proposition \ref{prop:square}
$T_{ab}\in \DP\cup -\DP$.\N

The singular case must be treated separately because of some minor
subtleties.
\begin{lem}
{\em (a)} $(T{}_2\!\times_2 T)_{ab}=0 \Longleftrightarrow T_a{}^b$ is a
singular null
cone preserving map $\Longleftrightarrow T_{ab}=s_a k_b$ where $k_b$ is null.

{\em (b)} $(T{}_2\!\times_2 T)_{ab}=(T{}_1\!\times_1 T)_{ab}=0
\Longleftrightarrow T_a{}^b$ is a singular null-cone bi-preserving map
$\Longleftrightarrow T_{ab}=n_a k_b$ where $n_a$ and $k_b$ are null
and $T_{ab}\in \DP\cup -\DP$.
\label{lem:sing}
\end{lem}
\P From Lemma \ref{lem:TxT=fg} we know that $(T{}_2\!\times_2 T)_{ab}=0$
if and only if the map defined by $T_a{}^b$ preserves the null cone, and by
Lemma \ref{lem:22=11} this map must be singular. Thus, from
Proposition \ref{prop:zerosquare} there exists a null $k_b$
such that $T_{ab}=s_a k_b$. This proves (a). Then, (b) follows from
Corollary \ref{cor:zerosquare} in a similar way.\N
\begin{coro}
The tensors in $\SS \setminus \NS$ (respectively in $-\SS \setminus -\NS$)
are proportional to involutory orthochronus (resp.\ time-reversal) Lorentz
transformations. The tensors in $\NS$ (resp.\ $-\NS$) define singular
orthochronus (resp.\ time-reversal) null-cone bi-preserving maps.
\label{cor:ssmaps}
\end{coro}
\P This follows at once from Proposition \ref{prop:txt},
Corollary \ref{cor:NxN}, Lemmas \ref{lem:TxT=fg}, \ref{lem:sing},
and the fact that if $T_a{}^b$ is involutory then by Lemma \ref{lem:22=11}
it must coincide with an involutory Lorentz transformation $T_a{}^b=L_a{}^b$,
which are symmetric $L_{ab}=L_{ba}$ \cite{PR}.\N

If a null-cone preserving map is non-symmetric (ergo not proportional to an
involutory Lorentz transformation if non-singular), then it can be
divided into its symmetric and anti-symmetric parts:
$$
T_{ab}\equiv S_{ab}+F_{ab},\hspace{1cm} S_{ab}\equiv T_{(ab)},\hspace{5mm}
F_{ab}\equiv T_{[ab]}\, .
$$
Notice that, by definition, if $T_{ab}$ is proportional to an
involutory Lorentz transformation then $T_{[ab]}=0$ and (up to sign)
$T_{(ab)}\in \DP$ (later we shall prove that, in fact, $T_{(ab)}\in
\SS$, see Theorem \ref{theo:fund}).
The general characterization is (see \cite{CSJ,CSJ2,PR} for $N=4$):
\begin{lem}
The symmetric and
antisymmetric parts of $T_a{}^b$ satisfy
\be
(S\times S)_{ab}+(F\times F)_{ab}=fg_{ab}, \hspace{1cm} (S\times F)_{(ab)}=0
\label{SF}
\ee
if and only if $T_a{}^b$ defines a null cone bi-preserving map.
Furthermore, $S_{ab}\in \DP\cup -\DP$.
\label{lem:SF}
\end{lem}
\P If $(T{}_2\!\times_2 T)_{ab}=(T{}_1\!\times_1 T)_{ab}=fg_{ab}$ then
\b*
S_{ac}S_b{}^c+S_{ac}F_b{}^c+F_{ac}S_b{}^c+F_{ac}F_b{}^c=fg_{ab} \\
S_{ac}S_b{}^c-S_{ac}F_b{}^c-F_{ac}S_b{}^c+F_{ac}F_b{}^c=fg_{ab}
\e*
and by adding and substracting these two equations the expresions (\ref{SF})
are obtained. Moreover, due to Lemmas \ref{lem:22=11} and \ref{lem:sing} (b)
we know that $T_{ab} \in \DP\cup -\DP$.
Then, from Lemma \ref{lem:perm} it follows that $S_{ab}\in \DP\cup -\DP$.\N

Recall that, from elementary considerations, any eigenvector of a 2-form
$F_{ab}$ with non-zero eigenvalue must be null. If there is one such
eigenvector, then there are exactly two of them with non-zero eigenvalues of
opposite signs, and any other eigenvector must be spacelike. Thus, if there is
a timelike eigenvector then all eigenvectors have zero eigenvalue. The
possible
number of null eigenvectors for a 2-form is: (i) if $N=2$, there are
exactly two
of them with nonzero eigenvalues of opposite sign; (ii) if $N=3$,
there can be $0,1$ or $2$ null eigenvectors; if there are $2$ then both of
them
have non-zero eigenvalues of opposite sign. (iii) for $N>3$ and even, say
$N=2n$ ($n\geq 2$), there can be either $1,2,4,\dots ,2(n-1)=N-2$ null
eigenvectors (the only odd number in the list is $1$). (iv) for $N>3$ and odd,
say $N=2n+1$, there can be either $0,1,2,3,\dots ,2n-1=N-2$ null eigenvectors
(the only even numbers in the list are $0,2$). In all cases, if there is only
one null eigenvector its eigenvalue is zero. This case includes the null
2-forms.
\begin{lem}
If $T_a{}^b$ defines a null cone bi-preserving map then:

{\em (a)} every null eigenvector of its symmetric part $T_{(ab)}$ is also a
null
eigenvector of its antisymmetric part $T_{[ab]}$.

{\em (b)} every eigenvector with non-zero eigenvalue of $T_{[ab]}$ is also
a null
eigenvector of $T_{(ab)}$.

{\em (c)} In the singular case, $T_{ab}=k_a n_b$, and $k^b$ and $n^b$ (which
may coincide if $T_{[ab]}=0$) are the
null eigenvectors of both $T_{(ab)}$ and $T_{[ab]}$.

{\em (d)} every null eigenvector $k^a$ with zero eigenvalue of $T_{[ab]}$ is
either a null eigenvector of $T_{(ab)}$ or there is another independent null
eigenvector $n^b$ with vanishing eigenvalue of $T_{[ab]}$ such that the
timelike 2-plane generated by $\{\vec{k},\vec{n}\}$ contains two eigenvectors
of both $T_{(ab)}$ and $T_{[ab]}$, one of them spacelike the other timelike,
with opposite eigenvalues.

{\em (e)} every timelike eigenvector of its symmetric part $T_{(ab)}$
is either an eigenvector also of $T_{[ab]}$ or there are two null
vectors which are simultaneously eigenvectors of both $T_{(ab)}$ and
$T_{[ab]}$ with non-zero eigenvalues.

{\em (f)} if $T_{[ab]}$ has a timelike eigenvector then there is 
a common timelike eigenvector for $T_{[ab]}$ and $T_{(ab)}$.

{\em (g)} Furthermore, if $T_{[ab]}\neq 0$ then $T_{(ab)}\notin \SS\cup
-\SS$ (for $N\neq 2$).
\label{lem:SFnull}
\end{lem}
\P Let us start with the null eigenvectors.
 From equations (\ref{SF}) we get for any null vector $k^a$
\bea
(k^a S_{ac})(k^b S_{b}{}^c)+(k^a F_{ac})(k^b F_{b}{}^c)=0, \label{kk1}\\
(k^a S_{ac})(k^b F_{b}{}^c)=0 .\hspace{1cm}\label{kk2}
\eea
Thus, if $k^a$ is an eigenvector of $S_{ab}$ then by (\ref{kk1}) $k^a F_{ac}$
is null and obviosuly orthogonal to $k^c$ so that they must be proportional
$k^a F_{ac}\propto k_c$. This proves (a). If $k^a$ is an eigenvector of
$F_{ab}$
then by (\ref{kk1}) $k^a S_{ac}$ is null, and by (\ref{kk2}) it is orthogonal
to $k^a F_{ac}=\lambda k_c$. Hence, if $\lambda\neq 0$ then $k^a S_{ac}\propto
k_c$, which proves (b). The statement (c) for the singular case follows
immediately from Lemma \ref{lem:sing} (b).
It remains the case with $k^a F_{ac}=0$. In this case from (\ref{SF}) we get
$$
(k^a S_{a}{}^{c})F_{bc}=0, \hspace{1cm} (k^a S_{a}{}^{c})S_{bc}=fk_b
$$
so that $k^a S_{a}{}^{c}$ is also a null eigenvector of $F_{ab}$ with zero
eigenvalue. If $k^a S_{a}{}^{c}\equiv n^c$ and $k^c$ are not colinear, that
is $\bm{k} \wedge \bm{n}\neq 0$, then $n^b\pm\sqrt{f}k^b$ are eigenvectors
of $S_{ab}$ with eigenvalues $\pm\sqrt{f}$, respectively. From (c) we know
that $f\neq 0$, so obviously one of
these vectors is timelike and the other spacelike, and both of them are
eigenvectors with zero eigenvalue of $F_{ab}$. This proves (d).

Concerning timelike eigenvectors, let $\vec u$ be unit and such that
$S_{ab}u^b=\lambda u_{a}$. Contracting relations (\ref{SF}) with $\vec u$ we
get
\be
(u^aF_{ac})F_{b}{}^c=(f-\lambda^2)u_{b}\, , \hspace{1cm}
S_{b}{}^c(u^aF_{ac})=\lambda (u^aF_{ab})\, .\label{unamas}
\ee
Thus, either $p_{c}\equiv u^aF_{ab}$ vanishes or it is spacelike (for
it is orthogonal to $u^c$). In the latter case from (\ref{unamas}) we
have $(p\cdot p)=f-\lambda^2<0$ and the two null vectors
$\vec u\pm \vec p/|p\cdot p|^{1/2}$ are eigenvectors of $S_{ab}$ with
eigenvalue $\lambda$, and also eigenvectors of $F_{ab}$ with
eigenvalues $\pm |p\cdot p|^{1/2}$ respectively, proving (e). To
prove (f), let $\vec u$ be unit and such that $F_{ab}u^b=0$. Then,
from (\ref{SF}) it follows
\be
(u^aS_{ac})S_{b}{}^c=f u_{b}\, , \hspace{1cm}
F_{b}{}^c(u^aS_{ac})=0\, .\label{otramas}
\ee
 From Lemma \ref{lem:SF} we know that $S_{ab}\in \DP\cup -\DP$, and then
$v_{c}\equiv u^aS_{ac}$ is causal. In fact, contracting the first relation
in (\ref{otramas}) with $u^b$ we deduce $(v\cdot v)=f$, so that
$\vec v$ must be timelike, as otherwise $f$ would vanish which is
impossible due to (c) above. Then, using (\ref{otramas}) is easy to
check that the two vectors $\vec{v}\pm\sqrt{f}\vec u$ are
eigenvectors of both $S_{ab}$ and $F_{ab}$, with eigenvalues
$\pm\sqrt{f}$ respectively, one of them timelike and the other spacelike.

Finally, to prove (g), if $\epsilon S_{ab}$ is in $\SS$ for $\epsilon =1$
or $-1$, then by Proposition \ref{prop:txt} $(S\times S)_{ab}=h^2 g_{ab}$
so that from (\ref{SF}) we have $(F\times F)_{ab}=(f-h^2)g_{ab}$. But then
Corollary \ref{cor:f>0} and Lemma \ref{lem:F} imply that $F_{ab}=0$
unless $N=2$. \N

Thus, the maps preserving the
null cone have a symmetric part which is in $\DP$
and either in $\SS$ (if $F_{ab}=0$, see Theorem \ref{theo:fund}) or not
(if $F_{ab}\ne 0$), in the second case algebraically
determined by the antisymmetric part of the map and its
null eigenvalues. Hence, in order to classify all these maps we only need to
know the structure of tensors in $\DP_2$ (defined as the rank-2 tensors
in $\DP$) in relation with $\SS\subset \DP_2$ and
with the null eigenvectors. Curiously enough, this result is the analogue
to Lemma \ref{lem:bb1} but for rank-2 symmetric tensors ($\DP_2$ and $\SS$
playing the role analogous to causal and null future-pointing
vectors, respectively): we now show that all symmetric tensors in $\DP_2$
can be written as sums of terms in $\SS$. This means that the elements in
$\SS$ can be used to build up $\DP_2$, and a fortiori $\SE_n$.
Furthermore, each term of $\SS$ in the
sum is related in a precise way to the null eigenvectors of the tensor in
$\DP_2$. More precisely, we have:
\begin{theo}
In $N$ dimensions, any symmetric rank-2 tensor $S_{ab}\in \DP_2$
can be written
\be
S_{ab}=\sum_{p=1}^N T_{ab}\{\Omega_{[p]}\}
\label{bb2}
\ee
where $T_{ab}\{\Omega _{[p]}\}\in \SS$
are the superenergy tensors of \underline{simple} $p$-forms
$\Omega_{[p]}$, $p=1,...,N$
such that for $p>1$ they have
the structure $\Omega_{[p]}=\bm{k}^{1}\wedge \dots \wedge \bm{k}^{p}$
where $\bm{k}^{1},\dots ,\bm{k}^{p}$ are appropriate
{\em null} 1-forms. The number of tensors in the sum
(\ref{bb2}) and the structure of the $\Omega_{[p]}$
depend on the particular $S_{ab}$ as follows:
if $S_{ab}$ has $N-m\geq 1$ null eigenvectors $\vec{k}^1,\dots,\vec{k}^{N-m}$
then {\em at least} $T_{ab}\{\Omega_{[N-m]}\}$, with
$\Omega_{[N-m]}=\bm{k}^{1}\wedge \dots \wedge \bm{k}^{N-m}$,
must appear in the sum, and possibly terms with $p>N-m$.
If it has no null eigenvectors, then {\em at least}
$T_{ab}\{\Omega_{[1]}\}$ appears in the sum, and possibly terms with
$p>1$, and $\Omega_{[1]}$ is the timelike eigenvector of $S_{ab}$.
\label{theo:bb2}
\end{theo}
{\it Remark:} As already stated, the superenergy tensor of the dual of a
$p$-form ($p<N$) is identical with that of the $p$-form itself. Thus, in the sum
(\ref{bb2}) there are two superenergy tensors of 1-forms, namely
$T_{ab}\{\Omega_{[1]}\}$ and $T_{ab}\{\Omega_{\stackrel{*}{[1]}}\}
=T_{ab}\{\Omega_{[N-1]}\}$, but the first one is the superenergy tensor of
a {\em causal} 1-form and the second of a {\em spacelike} 1-form. This is
an essential difference. Similar remarks apply to the 2-forms $\Omega_{[2]}$
and $\Omega_{[N-2]}$, and so on.
The choice of simple $p$-forms
taken in Theorem \ref{theo:bb2} is such that
$(-1)^{p-1}(\Omega_{[p]} \cdot \Omega_{[p]})> 0$ for $p=2,\dots ,N-1$,
and $\Omega_{[1]}$ is causal.

\noindent
\P Recall that for a symmetric tensor $S_{ab}$ any two eigenvectors with
different eigenvalues must be orthogonal. Then, any two linearly independent
{\em null} eigenvectors of a symmetric tensor must have the same eigenvalue.

We divide up in cases depending on the number of null eigenvectors of
$S_a{}^b$.
Suppose that $S_a{}^b$ has $N$ linearly independent null eigenvectors.
All their eigenvalues must be equal to some constant, say $\alpha$,
and $\alpha \ge 0$ as $S_{ab}\in \DP$. The $N$
null eigenvectors span all tangent vectors so we get $S_{ab}=\alpha g_{ab}=
T_{ab}\{\sqrt{2\alpha}\eta_{[N]}\}$ where $\bm\eta$ is the volume
$N$-form.

Suppose now that the Theorem is proven for the case with $(N-m)+1$
linearly independent null eigenvectors and assume that $S_a{}^b$ has
$N-m\ge 2$ linearly independent null eigenvectors,
$\vec{k}^{(1)},\dots ,\vec{k}^{(N-m)}$ say, all with eigenvalue $\beta$.
With ${}^*(\bm{k}^{1}\wedge \dots \wedge \bm{k}^{N-m})=
\bm{r}^{1}\wedge \dots \wedge \bm{r}^{m}$, all $\vec{r}^{(i)}$ must
be spacelike as they are orthogonal to all $\vec{k}^{(j)}$ and $N-m\ge 2$
(a vector which is orthogonal to two null vectors must be spacelike).
We have $S^{ab}k_a^{(j)} r_b^{(i)}=\beta \vec{k}^{(j)} \cdot\vec{r}^{(i)}=0$
so $S^{ab} r_b^{(i)}$ is orthogonal to all $\vec{k}^{(j)}$
and hence $S^{ab} r_b^{(i)}\in Span\{
\vec{r}^{(1)},\dots ,\vec{r}^{(m)}\}\equiv V_{(m)}$. Therefore $S_{ab}$ is a symmetric map
on $V_{(m)}$ which is a Euclidean space.
There are then $m$ orthonormal ($ y^{(i)} \cdot y^{(j)}=-\delta^{ij}$)
 eigenvectors $\vec{y}^{(1)},\dots ,\vec{y}^{(m)}$
to $S_{ab}$ in $V_{(m)}$ with corresponding
eigenvalues $ \lambda_1,\dots ,\lambda_m$. We can assume
$\lambda_1 \ge \lambda_i $ for $i=2,\dots , m$. Now, by Proposition
\ref{prop:eigen} and Corollary \ref{cor:Nevec},
$T_{ab}\{\Omega_{[N-m]}\}$, where
$\Omega_{[N-m]}=\bm{k}^{1}\wedge \dots \wedge \bm{k}^{N-m}$,
has also the null eigenvectors $\vec{k}^{(1)},\dots ,\vec{k}^{(N-m)}$
with some eigenvalue $\gamma> 0$ and it has the spacelike
eigenvectors $\vec{y}^{(1)},\dots ,\vec{y}^{(m)}$ with eigenvalues
$-\gamma$. Define $\tau_{ab}=S_{ab}+(\lambda_1 -\beta )
T_{ab}\{\Omega_{[N-m]}\}/2\gamma$. Then $\tau_a{}^b k_b^{(j)}=
(\beta +\lambda_1 )k_a^{(j)} /2$ and  $\tau_a{}^b y_b^{(1)}=
(\beta +\lambda_1 )y_a^{(1)} /2$. Take $u_a\in V_{(m)}^\perp$
with $u_a u^a=1$ and let $\tilde k_a =u_a +y_a^{(1)}$.
Then $\tilde k_a \not\in V_{(m)}^\perp$,
$\tilde k_a\tilde k^a=0$ and $\tau_a{}^b \tilde k_b=
(\beta +\lambda_1 )\tilde k_a /2$. Hence $\tau_{ab}$ has
$(N-m)+1$ linearly independent null eigenvectors
$\vec{k}^{(1)},\dots ,\vec{k}^{(N-m)},\tilde k$.
Note that
$S_a{}^b (u_b \pm y_b^{(i)})=\beta u_b \pm\lambda_i y_b^{(i)}$ are
future-pointing since $S_{ab}\in \DP$, and therefore
$\beta \ge |\lambda_i| $ for all $i$.
To show that $\tau_{ab}\in \DP$, then use that an arbitrary future-pointing
null vector $N_a$ can be written $N_a=K_a +Y_a$ where
$K_a=a_1 k_a^{(1)}+\dots +a_{N-m}k_a^{(N-m)}$ and
$Y_a=b_1 y_a^{(1)}+\dots +b_{m}y_a^{(m)}$, and where
$K_aK^a-b_1^2-\dots -b_m^2 =0$. Then $\tau_a{}^b N_b=
(\beta +\lambda_1 )K_a /2+\lambda_1 b_1 y_a^{(1)}+\dots\lambda_m b_m y_a^{(m)}
+(\beta -\lambda_1 )Y_a /2$ which has the squared length
$(\tau_a{}^b N_b)( \tau^{bc} N_c )= (\lambda_1 -\lambda_2)(\beta+
\lambda_2 )b_2^2+\dots +(\lambda_1 -\lambda_m)(\beta+
\lambda_m )b_m^2\ge 0$.
Thus, $S_{ab}=(\beta -\lambda_1 )
T_{ab}\{\Omega_{[N-m]}\}/2\gamma+\tau_{ab}=
T_{ab}\{\sqrt{(\beta -\lambda_1 )/2\gamma}\Omega_{[N-m]}\}+\tau_{ab}$
has the form required by the induction hypothesis so the statement of the
theorem holds
for the cases with at least 2 linearly independent null eigenvectors.

Next consider the case with precisely one null eigenvector
$\vec{k}$, with eigenvalue $\beta$. Take a set
$\{ \tilde{r}^{(1)},\dots ,\tilde{r}^{(N-2)}\}$ of linearly independent
spacelike vectors, all orthogonal to $\vec{k}$. Again we have
$S^{ab}k_a\tilde{r}_b^{(i)}=\beta k\cdot\tilde{r}^{(i)}=0$ so
$S^{ab}\tilde{r}_b^{(i)}$ is orthogonal to $\vec{k}$.
For those $i$ that $S_a{}^b\tilde{r}_b^{(i)}=\mu_ik_a$ for some real
number $\mu_i$, define $r_a^{(i)}=\tilde r_a^{(i)}-\mu_i k_a/\beta$ so
$S_a{}^b r_b^i=0$, $r_a^{(i)}$ is spacelike, and
$r_a^{(i)}k^a=0$. For those $i$ that $S_a{}^b\tilde{r}_b^{(i)}$ is
already spacelike, let $r_a^{(i)}=\tilde r_a^{(i)}$.
Let $\vec{n}$ be the other future-pointing null vector
orthogonal to $V_{(N-2)}=Span\{ \vec{r}^{(1)},\dots ,\vec{r}^{(N-2)}\}$
and normalised by $k_an^a=1$. As $S^{ab}r_b^{(i)}\in V_{(N-2)}$
we have $n_a(S^{ab}r_b^{(i)})=0$ which means that $S^{ab}n_b$ is orthogonal to
all $\vec{r}^{(i)}$. Thus, $S^{ab}n_b=\beta n^a+\gamma k^a$ with $\gamma >0$.
Define $\tau_{ab}=S_{ab} -\gamma k_ak_b$. Then $\tau_{ab}k^b=\beta k_a$
and $\tau_{ab}n^b=\beta n_a$ so $\vec{k}$ and $\vec{n}$ are two linearly
independent null eigenvectors of $\tau_{ab}$. To show $\tau_{ab}\in \DP$
we use as above that in $V_{(N-2)}$ there is an orthonormal basis of
eigenvectors $\{ \vec{y}^{(1)},\dots,\vec{y}^{(N-2)}\}$
where $\tau_a{}^{b}y_b^{(i)}=\lambda_i y_a^{(i)}$. For any $c>0$, $\ell_a=
ck_a+n_a/2c +y_a^{(i)}$ is future-pointing and null, and therefore
$0\le (S_{ab}\ell^b)(S^{ac}\ell_c
)=\beta^2+\beta\gamma/2c^2-\lambda_i^2$.
As $c$ can be taken arbitrary large we get $\beta^2\ge\lambda_i^2$.
An arbitrary future-pointing null vector can be written
$N_a=a_1k_a+a_2n_a+b_1y_a^{(1)}+\dots +b_{N-2}y_a^{(N-2)}$ where
$2a_1a_2-b_1^2-\dots -b_{N-2}^2=0$. We find
$(\tau_{ab}N^b)(\tau^{ac}N_c )=b_1^2(\beta^2-\lambda_1^2)+\dots
+b_{N-2}^2(\beta^2-\lambda_{N-2}^2)\ge 0$ and conclude that
$\tau_{ab}\in \DP$. Thus, $S_{ab}=\gamma k_ak_b+\tau_{ab}=
T_{ab}\{\sqrt{\gamma}k_{[1]}\}+\tau_{ab}$ has the required form
which proves the case with one null eigenvector.

Finally we consider the case with no null eigenvector.
If there exist null vectors $\vec{k}$ and $\vec{n}$
such that $S_{ab}k^a=n_b$ then, as $S_{ab}\in \DP$, $S_{ab}n^a=\beta k_b$
for some $\beta >0$. Then $\vec{n}+\sqrt\beta\vec{k}$ is a timelike 
eigenvector which normalised we denote by $\vec{u}$.
Otherwise, if all null vectors are mapped on
timelike vectors then again \cite{BR} $S_{ab}$ has a (unit) timelike 
eigenvector $\vec{u}$. Thus we have a unit timelike eigenvector $\vec{u}$
with eigenvalue $\lambda_0$,
and on $\{\vec{u}\}^\perp$ there is an ON-basis 
$\{ \vec{y}^{(1)},\dots,\vec{y}^{(N-1)}\}$ of eigenvectors with eigenvalues
$\lambda_1,\dots\lambda_{N-1}$. As $\vec{u}+\vec{y}^{(i)}$ is future-pointing
and null $S_a{}^{b}(u_b+y_b^{(i)})$ must be future-pointing which implies
$\lambda_0\ge |\lambda_i|$ for all $i=1,\dots ,N-1$. Assume
$\lambda_1 \ge \lambda_i $ for $i=2,\dots , N-1$ and define
$\tau_{ab}=S_{ab} -(\lambda_0-\lambda_1)(u_a u_b-{1\over 2}g_{ab})$. 
Then $\tau_a{}^b (u_b\pm y_b^{(1)})=
{1\over 2}(\lambda_0+\lambda_1)(u_a\pm y_a^{(1)})$ so $\tau_{ab}$
has {\it two} null eigenvectors.
To show that $\tau_{ab}\in \DP$,
let $c_1^2+\dots +c_{N-1}^2=1$; then $N_a(c_1,\dots,c_{N-1})=
(u_a+ c_1 y_a^{(1)}+\dots c_{N-1} y_a^{(N-1)})$ is
proportional to an arbitrary future-pointing null vector. One finds
$(\tau_{ab}N^b)(\tau^{ac}N_c )=
\Sigma_i c_i^2(\lambda_1-\lambda_i)(\lambda_0+\lambda_i) \ge 0$, so $S_{ab}=
T_{ab}\{\sqrt{\lambda_0-\lambda_1}u_{[1]}\}+\tau_{ab}$ has the right properties and 
this finishes the proof.
\N

{\it Remarks:} Recall that by Lemma \ref{lem:bb1} a future-pointing causal
vector can be written as a sum of two future-pointing null vectors in
infinitely many ways. In the same manner, a symmetric $S_{ab}\in \DP_2$
can be expressed as a sum of $N$ elements of
$\SS$ in many ways. As an example, let
$\left\{\bm{e}_{a}\right\}$ be an orthonormal
basis. Then, by (\ref{sep}), one easily find relations such as
\bea
T_{ab}\{(\bm{e}_{0})_{[1]}\}+T_{ab}\{(\bm{e}_{1})_{[1]}\}=
T_{ab}\{(\bm{e}_{0}\wedge\bm{e}_{2})_{[2]}\}+
T_{ab}\{(\bm{e}_{1}\wedge\bm{e}_{2})_{[2]}\},\nonumber\\
\alpha T_{ab}\{(\bm{e}_{1})_{[1]}\}+\beta T_{ab}\{(\bm{e}_{2})_{[1]}\}=
\frac{\beta}{2} g_{ab}+(\alpha -\beta)T_{ab}\{(\bm{e}_{1})_{[1]}\}+
\beta T_{ab}\{(\bm{e}_{1}\wedge\bm{e}_{2})_{[2]}\}. \label{util}
\eea
In Theorem \ref{theo:bb2} however, we construct the representation of
$S_{ab}\in \DP_2$
in a canonical way in which the simple $p$-forms $\Omega_{[p]}$
are constructed from the null eigenvectors of $S_{ab}$.

\section{Non-symmetric null-cone preserving maps}
We are now prepared to present the classification of
the general conformally non-involutory null-cone preserving maps, which follows
directly from the Theorem \ref{theo:bb2} and the Lemma \ref{lem:SFnull}.
Given that the results are elementary but the number of different
cases is increasing with the dimension $N$, we will restrict
ourselves to the low-dimension cases in full, but this will show the
way one has to follow as well as the general ideas which serve
for a general $N$. As the singular case has been
already solved, in this section we only deal with the
non-singular conformally non-involutory maps, so that $T_{[ab]}\neq 0$.
The conformally involutory ones are left for the next section.

{\bf Case} $N=2$. The simplest case is a 2-dimensional Lorentzian
manifold. In this case there are only two independent null directions,
say $\bm{\ell}$ and $\bm{k}$, and we can always write
$T_{[ab]}=(\bm{\ell}\wedge \bm{k})_{ab}=\mu \eta_{ab}$. Both $\vec{\ell}$ and
$\vec{k}$ are null eigenvectors of $T_{[ab]}$ with non-zero
eigenvalue, and then due to Lemma \ref{lem:SFnull} (b), they are also
null eigenvectors of $T_{(ab)}$. Using then Theorem \ref{theo:bb2} the
only possibility is that $T_{(ab)}=\alpha g_{ab}$. Thus, we have
\begin{coro}
In $N=2$, the maps proportional to non-involutory Lorentz transformations are
given by $T_{ab}=\alpha g_{ab}+\mu \eta_{ab}$ with arbitrary
$\alpha$ and $\mu$ such that $\alpha^2-\mu^2\neq 0$. They are
proper (resp. improper) if $\alpha^2-\mu^2 > 0$ (resp. $<0$), and orthochronus
(resp. time-reversal) if $\alpha >|\mu|$ (resp. $\alpha<-|\mu|$).
\end{coro}
Notice that in this particular case, an arbitrary 2-form
$\mu \eta_{ab}$ defines an improper null cone bi-preserving map.
This is the only possibility in which a 2-form can preserve the null
cone, and it appears as an exceptional case as follows from Corollary \ref{cor:f>0}
and Lemmas \ref{lem:TxT=fg} and \ref{lem:22=11}.

Before we proceed with the non-trivial cases $N>2$, we need some
simple lemmas. From Corollary \ref{cor:Nevec} we know that if
$T_{ab}\in \SS\setminus \NS$ then the tangent space can be decomposed
as $T_{x}(V_{N})=E^+\oplus E^-$ where $E^+$ is $p$-dimensional, $E^-$
is $(N-p)$-dimensional, and both $E^{\pm}$ are eigensubspaces of
$T_{ab}$ with opposite eigenvalues.

\begin{lem}
If $F_{[2]}$ is a simple 2-form and $T_{ab}\in \SS\setminus \NS$, then
$(T\times F)_{(ab)}=0$ if and only if $F_{ab}$ lies entirely in either
$\Lambda_{2}(E^+)$ or $\Lambda_{2}(E^-)$.
\label{lem:E+-}
\end{lem}
\P $F=\bm{\theta}_{1}\wedge \bm{\theta}_{2}$ for some 1-forms
$\bm{\theta}_{1}$ and $\bm{\theta}_{2}$. Obviously
$\vec{\theta}_{1}=\vec{\theta}_{1}^+ +\vec{\theta}_{1}^-$,
$\vec{\theta}_{2}=\vec{\theta}_{2}^+ +\vec{\theta}_{2}^-$ with
$\vec{\theta}_{1}^+,\vec{\theta}_{2}^+\in E^+$ and
$\vec{\theta}_{1}^-,\vec{\theta}_{2}^-\in E^-$. A straightforward
computation gives then
$$
(T\times F)_{(ab)}=2\lambda \left[(\theta_{1}^-\otimes
\theta_{2}^+)_{(ab)}-(\theta_{1}^+\otimes \theta_{2}^-)_{(ab)}\right]
$$
where $\lambda$ is the eigenvalue for $E^+$. Then, the condition
$(T\times F)_{(ab)}=0$ holds if and only if either
$\vec{\theta}_{1}^+=\vec{\theta}_{2}^+=0$ or
$\vec{\theta}_{1}^-=\vec{\theta}_{2}^-=0$.\N

Similarly, from Corollary \ref{cor:Nevec} if
$T_{ab}\left\{\Omega_{[p]}\right\}\in \NS$, there is an $(N-1)$-dimensional
subspace $E^0$ of eigenvectors with zero eigenvalue generated by
$\{\vec k,\vec{\omega}_{2},\dots ,\vec{\omega}_{N-1}\}$, where $\vec
k$ is the canonical null direction of the null $\Omega_{[p]}$.
\begin{lem}
If $F_{[2]}$ is a simple 2-form and $T_{ab}\in \NS$, then
$(T\times F)_{(ab)}=0\Longleftrightarrow (T\times F)_{ab}=0\Longleftrightarrow
F_{ab}$ lies entirely in $\Lambda_{2}(E^0)$.
\label{lem:E0}
\end{lem}
\P Set $F=\bm{\theta}_{1}\wedge \bm{\theta}_{2}$ as before and
choose $\vec n$ null, independent of $\bm{k}$, and orthogonal to all
$\{\vec{\omega}_{2},\dots ,\vec{\omega}_{N-1}\}$. Obviously
$\vec{\theta}_{1}=\vec{\theta}_{1}^0 +C_{1}\vec n$,
$\vec{\theta}_{2}=\vec{\theta}_{2}^0 +C_{2}\vec n$ with
$\vec{\theta}_{1}^0,\vec{\theta}_{2}^0\in E^0$. As
$T_{ab}=fk_{a}k_{b}$, we have
$$
(T\times F)_{ab}=f(k\cdot n) k_{a}(C_{1}\theta_{2}^+
-C_{2}\theta_{1}^+)_{b}
$$
and given that $\bm{\theta}_{1}$ and $\bm{\theta}_{2}$ are linearly
independent, the vanishing of this (or of its symmetric part) gives
$C_{1}=C_{2}=0$, and conversely.\N

The notation of Lemma \ref{lem:SFnull} for $T_{(ab)}=S_{ab}$
and $T_{[ab]}=F_{ab}$ is used in the remaining of this section.

\bigskip
{\bf Case} $N=3$. There are three possibilities, as $F_{ab}$ can have
0,1, or 2 null eigenvectors.

\noindent
(a) If $F_{ab}$ has no null eigenvector, then it is proportional to
the dual of a unit timelike vector $\vec u$, i.e. $F_{[2]}=\mu
u_{\stackrel{*}{[2]}}$. Due to Lemma \ref{lem:SFnull} (a), $S_{ab}$
has no null eigenvectors, and due to Lemma \ref{lem:SFnull} (f), $\vec
u$ is timelike eigenvector also of $S_{ab}$. Thus, Theorem
\ref{theo:bb2} allows us to write
$$
T_{ab}=\beta T_{ab}\left\{u_{[1]}\right\}+
\gamma T_{ab}\left\{\Omega_{[2]}\right\}+\alpha g_{ab}+ F_{ab}.
$$
Using Lemma \ref{lem:E+-} one has $(S\times F)_{(ab)}=
\gamma\left(T_{ab}\left\{\Omega_{[2]}\right\}\times F\right)_{(ab)}$
and here the term in brackets is non-vanishing due again to Lemma
\ref{lem:E+-}. Thus, the second condition (\ref{SF}) implies $\gamma
=0$. With this, it is easily checked that the first condition in
(\ref{SF}) leads to $\mu^2=2\alpha\beta$. Thus, we obtain
\be
T_{ab}=\beta T_{ab}\left\{u_{[1]}\right\}+\alpha g_{ab}\pm
\sqrt{2\alpha\beta} \, \left(u_{\stackrel{*}{[2]}}\right){}_{ab} \, .
\label{3a}
\ee
These maps are proper and orthochronus if $\alpha >0$, and improper
and time-reversal if $\alpha <0$.

\noindent
(b) If $F_{ab}$ has one null eigenvector $\vec k$, then $F_{[2]}$ is
null and can be written $F=\mu \bm{k}\wedge \bm{p}$ with $(k\cdot p)=0$.
Lemma \ref{lem:SFnull} (d) implies that $\vec k$ is also a null
eigenvector of $S_{ab}$, and this is unique for $S_{ab}$ due to Lemma
\ref{lem:SFnull} (a). So, again Theorem \ref{theo:bb2} tells us that
$$
T_{ab}=\beta T_{ab}\left\{k_{[1]}\right\}+
\gamma T_{ab}\left\{\Omega_{[2]}\right\}+\alpha g_{ab}+ F_{ab}.
$$
Analogously to case (a) above, Lemmas \ref{lem:E+-} and \ref{lem:E0}
lead to $\gamma =0$, and the first relation in
(\ref{SF}) gives again $\mu^2=2\alpha\beta$. Hence
\be
T_{ab}=\beta T_{ab}\left\{k_{[1]}\right\}+\alpha g_{ab}\pm
\sqrt{2\alpha\beta} \, \left(k_{\stackrel{*}{[2]}}\right){}_{ab} \, .
\label{3b}
\ee
Notice that this can be considered a limit case of (\ref{3a}) when
$\vec u$ becomes null.

\noindent
(c) If $F_{ab}$ has two independent null eigenvectors $\vec k$ and $\vec
n$, then they necessarily have non-zero eigenvalues, and by
Lemma \ref{lem:SFnull} (b) they are also eigenvectors of $S_{ab}$,
which cannot have more null eigenvectors due to Lemma
\ref{lem:SFnull} (a). Thus, by Theorem \ref{theo:bb2}
$$
T_{ab}=\beta T_{ab}\left\{(k\wedge n)_{[2]}\right\}+\alpha g_{ab}+
\mu (k\wedge n)_{ab}.
$$
The computation of (\ref{SF}) leads now simply to $\mu^2=2\alpha\beta$.
In summary,
$$
T_{ab}=\beta T_{ab}\left\{(k\wedge n)_{[2]}\right\}+\alpha g_{ab}\pm
\sqrt{2\alpha\beta} \, (k\wedge n)_{ab} \, .
$$
Observe that this case can be rewritten as
\be
T_{ab}=\beta T_{ab}\left\{p_{[1]}\right\}+\alpha g_{ab}\pm
\sqrt{2\alpha\beta} \, \left(p_{\stackrel{*}{[2]}}\right){}_{ab} \, .
\label{3c}
\ee
where $\vec p$ is spacelike and defined by $\bm{p}\equiv
*(\bm{k}\wedge \bm{n})$. Hence, the combination of
(\ref{3a}-\ref{3c}) proves the following
\begin{coro}
In $N=3$, the maps proportional to non-involutory Lorentz
transformations are given by
$$
T_{ab}=\beta T_{ab}\left\{\S_{[1]}\right\}+\alpha g_{ab}\pm
\sqrt{2\alpha\beta} \, \left(\S_{\stackrel{*}{[2]}}\right){}_{ab}
$$
where $\alpha$ and $\beta$ are arbitrary with $\alpha \beta >0$ and
$\S_{[1]}$ is any 1-form. These maps leave none, one or two null
directions invariant if $\S_{[1]}$ is time-, light-, or space-like,
respectively.\N
\end{coro}

\bigskip
{\bf Case} $N=4$. Now there are just two possibilities: either $F_{ab}$
has one or two null eigenvectors.

\noindent
(a) If $F_{ab}$ has one null eigenvector $\vec k$, then $F_{[2]}$ is
null, $F=\mu \bm{k}\wedge \bm{p}$ with $(k\cdot p)=0$. Due to Lemma
\ref{lem:SFnull} (d) and (a) this is also the unique null eigenvector
of $S_{ab}$ so that from Theorem \ref{theo:bb2}
$$
T_{ab}=\beta T_{ab}\left\{k_{[1]}\right\}+
\delta T_{ab}\left\{\Omega_{[2]}\right\}+
\gamma T_{ab}\left\{\Omega_{[3]}\right\}+\alpha g_{ab}+ F_{ab}.
$$
with $\Omega_{[2]}$ and $\Omega_{[3]}$ having the form
$\bm{k}\wedge \bm{n}$ and $\bm{k}\wedge \bm{n}\wedge \bm{\ell}$,
respectively, for null $\bm{n}$ and $\bm{\ell}$.
Lemmas \ref{lem:E+-} and \ref{lem:E0}
imply that the second equation in (\ref{SF}) reads
$$
\delta\left(T\left\{\Omega_{[2]}\right\}\times F\right)_{(ab)}+
\gamma\left(T\left\{\Omega_{[3]}\right\}\times F\right)_{(ab)} =0
$$
which, as $\vec p$ cannot be linear combination of $\vec k$ and $\vec
n$, becomes
$$
-(\delta \lambda +\gamma \tilde{\lambda})k_{(a}p_{b)}+
\gamma k_{(a}T_{b)c}\left\{\Omega_{[3]}\right\}p^c=0
$$
where $\lambda$ and $\tilde\lambda$ are positive as they are
proportional to $-(\Omega_{[2]}\cdot \Omega_{[2]})$ and
$(\Omega_{[3]}\cdot \Omega_{[3]})$, respectively. The above
expression can only be satisfied with non-negative $\delta\gamma \geq 0$
if $\delta=0$ and
$T_{bc}\left\{\Omega_{[3]}\right\}p^c=\tilde\lambda p_{b}$. This also
implies that $\bm{p}\wedge (\bm{k}\wedge \bm{n}\wedge \bm{\ell})=0$
and we can write
$$
T_{ab}=\beta T_{ab}\left\{k_{[1]}\right\}+
\gamma T_{ab}\left\{(k\wedge n\wedge \ell)_{[3]}\right\}
+\alpha g_{ab}+ F_{ab}.
$$
The remaining condition in (\ref{SF}) implies in particular that $\alpha\gamma
=0$, so that two possibilities arise (assuming that $\vec p$ is unit):
$\alpha =0$ and then $\mu^2=\beta\gamma$; or $\gamma =0$ and
$\mu^2=2\alpha\beta$. In summary, by setting
$\bm{q}\equiv *(\bm{k}\wedge \bm{n}\wedge \bm{p})$, we have
\bea
T_{ab}=\beta T_{ab}\left\{k_{[1]}\right\}+
\gamma T_{ab}\left\{q_{[1]}\right\}
\pm \sqrt{\beta\gamma} (k\wedge p)_{ab}, \label{4a1}\\
T_{ab}=\beta T_{ab}\left\{k_{[1]}\right\}+\alpha g_{ab}
\pm\sqrt{2\alpha\beta}(k\wedge p)_{ab}. \label{4a2}
\eea
Observe that in both cases one can replace
$T_{ab}\left\{k_{[1]}\right\}$ by $T_{ab}\left\{(k\wedge
p)_{[2]}\right\}$, because $F_{[2]}$ is null. Furthermore, the above expressions
(\ref{4a1}-\ref{4a2}) are valid for {\em arbitrary} $N$ so they are proportional
to Lorentz transformations in any $V_N$ (where $\vec q$ is just any spacelike
vector orthogonal to both $\vec k$ and $\vec p$). 

\noindent
(b) If $F_{ab}$ has two null eigenvectors $\vec k$ and $\vec n$,
then
$$
F_{[2]}=\mu_{1}(k\wedge n)_{[2]}+\mu_{2}(k\wedge n)_{\stackrel{*}{[2]}}
$$
with $\mu_{1}^2+\mu_{2}^2\neq 0$. If $\mu_1\neq 0$,
from Lemma \ref{lem:SFnull} (a), (b)
and (d), $\vec k$ and $\vec n$ are the two null eigenvectors of
$S_{ab}$ and we can write in principle, from Theorem \ref{theo:bb2},
$$
T_{ab}=\gamma T_{ab}\left\{\Omega_{[2]}\right\}+
\beta T_{ab}\left\{\Omega_{[3]}\right\}+\alpha g_{ab}+ F_{ab}.
$$
with $\Omega_{[2]}$ and $\Omega_{[3]}$ having the form
$\bm{k}\wedge \bm{n}$ and $\bm{k}\wedge \bm{n}\wedge \bm{\ell}$,
respectively, for null $\bm{\ell}$.
Lemmas \ref{lem:E+-} and \ref{lem:E0}
imply that the second equation in (\ref{SF}) leads to
$\beta\mu_{2}=0$. Solving these two possibilites we arrive at
\bea
T_{ab}=\gamma T_{ab}\left\{(k\wedge n)_{[2]}\right\}+
\alpha g_{ab}\pm \sqrt{2\alpha\gamma} \left[\cos\theta (k\wedge n)+
\sin\theta *(k\wedge n)\right]_{ab}, \label{4b1}\\
T_{ab}=2\alpha T_{ab}\left\{(k\wedge n)_{[2]}\right\}+
\beta T_{ab}\left\{(k\wedge n\wedge \ell)_{[3]}\right\}+\alpha g_{ab}
\pm\sqrt{2\alpha (\beta +2\alpha)}(k\wedge n)_{ab} \label{4b2}
\eea
where $\theta$ is arbitrary.

If $\mu_1=0$, there also arises the possibility given by
Lemma \ref{lem:SFnull} (d), (f) that $S_{ab}$ has a timelike eigenvector
$\vec u$ and a spacelike one $\vec p$ with
$\bm{k}\wedge \bm{n}=\bm{u}\wedge \bm{p}$, such that from Theorem
\ref{theo:bb2} one has in principle
$$
T_{ab}=\beta T_{ab}\left\{u_{[1]}\right\}+
\delta T_{ab}\left\{\Omega_{[2]}\right\}+
\gamma T_{ab}\left\{\Omega_{[3]}\right\}+\alpha g_{ab}+ F_{ab}.
$$
with $\Omega_{[2]}\wedge \bm{p}\neq 0$ and
$\Omega_{[3]}\equiv p_{\stackrel{*}{[3]}}$. However, $\vec u$ and
$\vec p$ have opposite eigenvalues due to Lemma \ref{lem:SFnull} (d), from
where we get $\alpha =0$. Then, from Lemma \ref{lem:E+-} and
the second equation in (\ref{SF}) it follows that $\delta =0$ too.
Finally, taking $\vec u$ and $\vec p$ unit,
the first relation in (\ref{SF}) leads to $\mu_2^2=2\beta\gamma$
so that
\be
T_{ab}=\beta T_{ab}\left\{u_{[1]}\right\}+
\gamma T_{ab}\left\{p_{[1]}\right\}\pm\sqrt{2\beta\gamma} (u\wedge p)_{ab}.
\label{4b3}
\ee
\begin{coro}
In $N=4$, the maps proportional to non-involutory Lorentz
transformations are given by (\ref{4a1}-\ref{4b3}). \N
\end{coro}
These results were obtained for the restricted case in
\cite{CSJ,CSJ2}, and in general in \cite{PR} using spinors. The case
given by (\ref{4b2}) may seem not included in the solution presented
in \cite{PR}, but this is apparent. In fact, one can rewrite (\ref{4b2})
by using the identity (\ref{util}) as
$$
2\alpha T_{ab}\left\{p_{[1]}\right\}+(2\alpha +\beta)
T_{ab}\left\{q_{[1]}\right\}\pm\sqrt{2\alpha (\beta +2\alpha)}(k\wedge n)_{ab}
$$
where $\bm{q}\equiv *(\bm{k}\wedge \bm{n}\wedge \bm{\ell})$ and
$\bm{p}\wedge \bm{q} \equiv *(\bm{k}\wedge \bm{n})$, and this last
form is certainly included in the cases given in \cite{PR}.

\bigskip
The number of possibilities and the
complexity of the equations increase with $N$, but the reasonings and
techniques are always simple and the same: application of
Lemmas \ref{lem:SFnull}, \ref{lem:E+-} and \ref{lem:E0} and Theorem
\ref{theo:bb2} to the equations (\ref{SF}). The details will be omitted
here but, as an illustrative example, we present the general solution
for arbitrary odd dimension $N=2n+1$.

{\bf Case} $N=2n+1$, ($n\geq 2$).
Let $\{\vec{e}_{0},\dots,\vec{e}_{2n}\}$ be an orthonormal basis. Then, the maps
proportional to non-involutory Lorentz transformations are in one of
the following cases:
\b*
{\rm (1)} \hspace{3mm}
T_{ab}=\beta_{1}T_{ab}\left\{v_{[1]}\right\}+\sum_{j=2}^n
\beta_{j}
T_{ab}\left\{(e_{2j-1}\wedge\dots \wedge e_{2n})_{\stackrel{*}{[2j-1]}}\right\}+
\alpha g_{ab}+\\
\pm\mu_{1}(v\wedge e_{2})_{ab}\pm\sum_{j=2}^n\mu_{j}(e_{2j-1}\wedge
e_{2j})_{ab} \hspace{2cm}
\e*
where $\alpha$, $\beta_{1},\dots,\beta_{n}$ are arbitrary, the
$\mu_{i}$ are given, for all $i=1,\dots ,n$ by
$$
\mu_{i}^2=(\beta_{1}+\dots +\beta_{i})(\beta_{i+1}+\dots
+\beta_{n}+2\alpha),
$$
and $\bm{v}$ is a causal 1-form equal to $\bm{e}_{0}$ if $T_{ab}$ leaves no
null direction invariant, and to $\bm{e}_{0}+\bm{e}_{1}$ if it leaves
exactly one null direction ($\vec v$) invariant.
\b*
{\rm (2)} \hspace{3mm}
T_{ab}=\beta_{1}T_{ab}\left\{(e_{0}\wedge e_{1})_{[2]}\right\}+\sum_{j=2}^n
\beta_{j}
T_{ab}\left\{(e_{2j-1}\wedge\dots \wedge e_{2n})_{\stackrel{*}{[2j-1]}}\right\}+
\alpha g_{ab}+\\
\pm\mu_{1}(e_{0}\wedge e_{1})_{ab}\pm\sum_{j=2}^n\mu_{j}(e_{2j-1}\wedge
e_{2j})_{ab} \hspace{2cm}
\e*
where now
$$
\mu_{1}^2=\beta_{1}(\beta_{2}+\dots +\beta_{n}+2\alpha),
$$
and for all $j=2,\dots,n$
$$
\mu_{j}^2=(\beta_{2}+\dots +\beta_{j})(\beta_{j+1}+\dots
+\beta_{n}+2\alpha -\beta_{1}).
$$
Generically, this leaves 2 null directions invariant, 3 if
$\beta_{1}=0$, and in general $2j+1$ null directions invariant if
$\beta_{1}=\dots=\beta_{j}=0$.

{\rm (3)} \hspace{3mm} Those cases which effectively reduce to low-dimensional
cases, such as for instance
$$
T_{ab}=2\alpha T_{ab}\left\{(e_{0}\wedge e_{1})_{[2]}\right\}+
\beta T_{ab}\left\{(e_{2})_{\stackrel{*}{[2n]}}\right\}+\alpha g_{ab}
\pm\sqrt{2\alpha (\beta +2\alpha)}(e_{0}\wedge e_{1})_{ab}
$$
which is the analogue of (\ref{4b2}) and
has two invariant null directions. And similarly for the
appropriate generalizations of (\ref{4a1}-\ref{4a2}) and (\ref{4b3}).

\section{Symmetric null-cone preserving maps and algebraic Rainich conditions}

We are now going to prove an important result: the converses of Proposition
\ref{prop:txt}, Lemma \ref{lem:sing} and Corollary \ref{cor:ssmaps} hold.
One can also intrinsically determine the rank $p$
of the $p$-form generating the tensor in $\SS$. More precisely
\begin{theo}
In $N$ dimensions, if $T_{ab}$ is symmetric and $(T\times T)_{ab}=fg_{ab}$
then:

{\em (a)} $f=0 \Longrightarrow T_{ab}\in \NS\cup -\NS$ and $T_{ab}=\beta
k_ak_b$
for a null $k_b$.

{\em (b)} $f\neq 0 \Longrightarrow  T_{ab}\in \SS\cup -\SS$ and,
for some $p\in\{1,\dots ,N\}$,
$\epsilon T_{a}{}^{a}=(2p-N)\sqrt{f}$ with $\epsilon=\pm 1$. Moreover,
$\epsilon T_{a}{}^{a}=(2p-N)\sqrt{f}\Longleftrightarrow
\epsilon T_{ab}=T_{ab}\{\Omega _{[p]}\}$
where $T_{ab}\{\Omega _{[p]}\}$ is the superenergy tensor of a simple
$p$-form $\Omega_{[p]}$ of the type used in Theorem \ref{theo:bb2}.
\label{theo:fund}
\end{theo}
\P By Corollary \ref{cor:f>0}, $T_{ab}$ symmetric and $(T\times
T)_{ab}=fg_{ab}$ implies that $f\ge 0$.
Thus $\sqrt{f}$ is well defined as the positive square root of $f$.
Lemmas \ref{lem:22=11} and \ref{lem:sing} give then $\epsilon T_{ab}\in \DP$.
Then by Theorem \ref{theo:bb2} and using $T_{ab}\{\Omega_{[N]}\}=\alpha g_{ab}$,
$\epsilon T_{ab}=T_{ab}\{\Omega_{[1]}\}+...+T_{ab}\{\Omega_{[N-1]}\}+
\alpha g_{ab}$. We have to verify that only one term or proportional terms can
remain if $(T\times T)_{ab}=fg_{ab}$. We have
$(T\times T)_{ab}=(\alpha g_{c(a}+T_{c(a}\{\Omega_{[1]}\}+...
+T_{c(a}\{\Omega_{[N-1]}\})(\alpha g_{b)}{}^{c}+T_{b)}{}^{c}\{\Omega_{[1]}\}
+...+T_{b)}{}^{c}\{\Omega_{[N-1]}\})$. This expression contains four
types of terms: $\alpha^{2}g_{ca}g_{b}{}^{c}$,
$\alpha g_{ca}T_{b}{}^{c}\{\Omega_{[p]}\}$,
$T_{ca}\{\Omega_{[p]}\}T_{b}{}^{c}\{\Omega_{[p]}\}$ and
$T_{ca}\{\Omega_{[p]}\}T_{b}{}^{c}\{\Omega_{[q]}\}$ with $p\neq q$.
 By Theorem \ref{th:dp} and Lemma \ref{lem:cont}
every term is in $\DP$. If $(T\times T)_{ab}=f g_{ab}$ then
$(T\times T)_{ab}k^{a}k^{b}=0$ for every null vector $k^{a}$.
Since each term is non-negative this means that every term
must be zero when contracted with $k^{a}k^{b}$. By Proposition \ref{prop:txt}
this is satisfied by $\alpha^{2}g_{ca}g_{b}{}^{c}$ and
$T_{ca}\{\Omega_{[p]}\}T_{b}{}^{c}\{\Omega_{[p]}\}$ since they are
proportional
to $g_{ab}$. Take then $\alpha g_{ca}T_{b}{}^{c}\{\Omega_{[p]}\}k^{a}k^{b}=0$.
This means that $\alpha T_{ab}\{\Omega_{[p]}\}k^{a}k^{b}=0$ which is
impossible
for every null vector $k^{a}$ unless $\alpha =0$ or $\Omega_{[p]}=0$.
Thus if $\alpha\neq 0$ then $\Omega_{[p]}=0$ for all $p=1,...,N-1$ so
$\epsilon T_{ab}=\alpha g_{ab}$ with $\alpha =\sqrt{f}$ and
$\epsilon T_{a}{}^{a}=N\sqrt{f}$.

If $\alpha=0$ we need to study the implications of
$T_{ca}\{\Omega_{[p]}\}T_{b}{}^{c}\{\Omega_{[q]}\}k^{a}k^{b}=0$ for every null
vector $k^{a}$. By Proposition \ref{prop:txt} and Lemma \ref{lem:TxT=fg},
$T_{ca}\{\Omega_{[p]}\}k^{a}$ and $T_{b}{}^{c}\{\Omega_{[q]}\}k^{b}$ are null
vectors and hence they must be parallel,
$T_{ca}\{\Omega_{[p]}\}k^{a}=\beta (k)T_{ca}\{\Omega_{[q]}\}k^{a}$.
Contraction with another null vector $n^a$ gives $\beta (n)=\beta (k)$
because of symmetry. Therefore, there is some constant $\beta$ such that
$T_{ca}\{\Omega_{[p]}\}k^{a}=\beta T_{ca}\{\Omega_{[q]}\}k^{a}$
for all null vectors, and extending by linear combinations to all vectors we
must have $T_{ca}\{\Omega_{[p]}\}=\beta T_{ca}\{\Omega_{[q]}\}$,
or that one of them is zero. In total, all non-zero $T_{ab}\{\Omega_{[p]}\}$,
$p=1,...,N-1$, should be proportional, but this is not possible because
as seen in the proof of Theorem \ref{theo:bb2} each of them has a strict
different number of null eigenvectors. Then
$\epsilon T_{ab}=T_{ab}\{\Omega_{[p]}\}$ for some $p$ and by (\ref{ssxss})
$(T\times T)_{ab}=[(\Omega \cdot \Omega )^{2}/(2\, p!)^{2}]\, g_{ab}$
so $f=[(\Omega \cdot \Omega )/(2\, p!)]^2$ and
$$
\epsilon T_{a}{}^{a}=T_{a}{}^{a}\{\Omega_{[p]}\}=\frac{(-1)^{p-1}}{(p-1)!}
(1-N/2p)(\Omega \cdot \Omega)=\frac{(2p-N)}{2\, p!}|\Omega \cdot \Omega |=
(2p-N)\sqrt{f}
$$
which, as $p=1,...,N-1$, gives the values $(2-N)\sqrt{f},...,(N-2)\sqrt{f}$.
Finally, the particular case (a), with $f=0$, follows immediately from
the above or directly from Lemma \ref{lem:sing} (b).\N

Combining Proposition \ref{prop:txt}, Lemma \ref{lem:sing}, Corollary
\ref{cor:ssmaps} and Theorem \ref{theo:fund}, we immediately
obtain the following important result.
\begin{theo}
In $N$ dimensions, if $T_{ab}$ is symmetric then
$(T\times T)_{ab}=fg_{ab} \Longleftrightarrow T_{ab}\in \SS\cup -\SS$.
This means that
the tensors in $\SS \setminus\NS$ (respectively in $-\SS \setminus-\NS$)
are precisely those proportional to involutory orthochronus
(resp.\ time-reversal) Lorentz transformations. The {\em symmetric}
tensors in $\NS$ (resp.\ $-\NS$) are exactly the symmetric singular
orthochronus (resp.\ time-reversal) null-cone bi-preserving maps.\N
\label{theo:fund2}
\end{theo}
Theorems \ref{theo:fund} and \ref{theo:fund2} provide a complete
characterization of the conformally involutory null-cone preserving maps.
Its classification also follows from the proof of Proposition \ref{prop:eigen}
and Corollary \ref{cor:Nevec}, for
we know that $p$ eigenvalues of $T_{ab}\{\Omega _{[p]}\}\in \SS$ are equal to
$(-1)^{p-1}(\Omega \cdot \Omega)/(2\, p!)$ while $(N-p)$ are equal to
$(-1)^{p}(\Omega \cdot \Omega)/(2\, p!)$. If an odd number of these are
negative, $T_{ab}\{\Omega _{[p]}\}$ is an improper null cone preserving
map, otherwise a proper one. If $\Omega_a$ is a spacelike 1-form,
then one eigenvalue is negative so $T_{ab}\{\Omega _{[1]}\}$ is improper
and can be interpreted as a reflection in the hyperplane orthogonal
to $\Omega_a$. If $\omega_a$ is a timelike 1-form,
then $(N-1)$ eigenvalues are negative and $T_{ab}\{\omega _{[1]}\}$
is proper in odd dimensions and improper in even dimensions.
It can be interpreted as a reflection in the line parallel to $\omega_a$.
For other $p$-forms one can develop the corresponding geometrical
interpretations.

Proposition \ref{prop:TxT} and Theorem \ref{theo:fund2} also imply:
\begin{coro}
If $N\leq 4$ then $\SE_2=\SS$, i.e. in 2, 3, and 4 dimensions, $\SE_2$
is precisely the set of symmetric tensors leaving invariant
the null cone with its time orientation. Thus, if $N\leq 4$, $\SE_2$ is
constituted by all tensors proportional to involutory orthochronus
Lorentz transformations plus the symmetric singular orthochronus null-cone
bi-preserving maps. $-\SE_2$ gives the time-reversal case.\N
\end{coro}
For $N\leq 3$ this is trivial. For $N=4$ this also means that the
energy-momentum of any Maxwell field is proportional to an involutory
orthochronus (and proper) Lorentz transformation, and coincides with
the energy-momentum of some (possibly another) Maxwell field
corresponding to a simple 2-form.
This is well known and related to the  duality rotations \cite{PR,PlPr}.
With $N=4$ and $p=2$ in Theorem \ref{theo:fund} we can state this as
\begin{coro}
In $4$ dimensions, a tensor $T_{ab}$ is (up to sign) {\em algebraically}
the energy-momentum tensor of a Maxwell field (a 2-form)
 if and only if $(T\times T)_{ab}=fg_{ab}$ and $T_a{}^a=0$.\N
\end{coro}
These are the classical algebraic Rainich conditions \cite{MW,R}
(see also \cite{Ha,Hl,ex,PR}).
They are necessary and sufficient conditions for a spacetime metric
to originate algebraically (via Einstein's equations) in a Maxwell field, i.e.
a way of determining the physics from the geometry.

By Theorem \ref{theo:fund} we can find generalisations to arbitrary
dimension and to many different physical fields. In order to show the
possibilities of our results, we can derive the following algebraic
Rainich conditions. For a scalar field (compare with the partial
results in \cite{Ku,Pen1,Pen2,Per} for $N=4$) we have
\begin{coro}
In $N$ dimensions, a tensor $T_{ab}$ is {\em algebraically}
the energy-momentum tensor of a minimally coupled massless scalar field
$\phi$ if and only if $(T\times T)_{ab}=fg_{ab}$ and $T_a{}^a =\beta
\sqrt{T_{ab}T^{ab}/N}$
where $\beta =\pm (N-2)$. Moreover, $d\phi$ is spacelike if $\beta =N-2$
and $T_a{}^a\ne 0$, timelike if $\beta =2-N$ and $T_a{}^a\ne 0$, and null
if $T_a{}^a =0=T_{ab}T^{ab}$.
\end{coro}
\P Recall that $T_{ab}=\nabla_a\phi\nabla_b\phi -
(1/2)(\nabla\phi\cdot\nabla\phi)g_{ab}$ which is exactly
$T_{ab}\{\nabla_{[1]}\phi \}\in \SS$ so $(T\times T)_{ab}=fg_{ab}$.
We get  $T_a{}^a =(2-N)
(\nabla\phi\cdot\nabla\phi)/2$ and
$T_{ab}T^{ab}=N(\nabla\phi\cdot\nabla\phi)^2
/4$ so $T_a{}^a =\pm (N-2) \sqrt{T_{ab}T^{ab}/N}$ with a plus sign if
$\nabla\phi\cdot\nabla\phi <0$ and a minus sign if
$\nabla\phi\cdot\nabla\phi >0$.
Conversely, if  $(T\times T)_{ab}=fg_{ab}$ then
$T_{ab}T^{ab}=Nf$. If $T_a{}^a =\pm (N-2) \sqrt{T_{ab}T^{ab}/N}=
\pm (N-2) \sqrt{f}$, then by Theorem \ref{theo:fund}
$T_{ab}$ is the superenergy tensor of a null 1-form in case $f=0$. If $f\ne 0$ and
$T_a{}^a = (2-N) \sqrt{f}$ then $T_{ab}$ is the superenergy tensor
of a timelike 1-form. If $f\ne 0$ and $T_a{}^a = (N-2) \sqrt{f}$
then $T_{ab}$ is the superenergy tensor of an $(N-1)$-form of the type
used in Theorem \ref{theo:bb2}
which is the same as the superenergy tensor of its dual spacelike 1-form.
\N

We can also generalize the algebraic Rainich conditions for a perfect fluid
as given by Coll and Ferrando \cite{CF} to the case of general $N$.
\begin{coro}
In $N$ dimensions, a tensor $T_{ab}$ is {\em algebraically}
the energy-momentum tensor of a perfect fluid satisfying the dominant
energy condition if and only if
\be
T_{ab}=\frac{\lambda}{2}g_{ab}+\mu T_{ab}\{v_{[1]}\} \label{pf}
\ee
where $\lambda ,\mu \geq 0$ and $v_b$ is timelike (so that $T_{ab}\{v_{[1]}\}$
is intrinsically characterized as a tensor in $\SS$ according to its trace,
see previous Corollary). The velocity vector of the fluid, its energy
density and pressure are given by $u^b=v^b/(v\cdot v)$,
$\rho =(\mu (v\cdot v)+\lambda)/2$ and $P=(\mu (v\cdot v)-\lambda)/2$,
respectively.
\end{coro}
\P Recall that a perfect fluid has the Segre type $\{1,(1\dots 1)\}$, so that
\be
T_{ab}=(\rho +P)u_a u_b -Pg_{ab} \label{pf2}
\ee
where $(u\cdot u)=1$. Thus, if (\ref{pf}) holds it is obvious that $T_{ab}$
takes the form (\ref{pf2}). Conversely, if (\ref{pf2}) holds, then
$T_{ab}-T_{ab}\{u_{[1]}\}$ has every null $k^b$ as eigenvector, as can be
trivially checked. Therefore, $T_{ab}-T_{ab}\{u_{[1]}\}$ is proportional to
the metric, and the proportionality factor is obtained from the $T_a{}^a$.\N

In fact, we can get the conditions as stated in \cite{CF} generalized for $N$
dimensions as follows. From (\ref{pf}) we get
$$
(T\times T)_{ab}=\lambda\mu T_{ab}\{v_{[1]}\}+
\frac{{\mu}^2(v\cdot v)^2+{\lambda}^2}{4}g_{ab}=
\lambda T_{ab}+\frac{{\mu}^2(v\cdot v)^2-{\lambda}^2}{4}g_{ab}=
\lambda T_{ab}+\rho P g_{ab}
$$
and also $N(T\times T)_a{}^a-(T_a^a)^2\geq 0$, $T_a^a\leq \frac{N}{2}\lambda$
and $T_{ab}w^aw^b\geq \lambda /2$ for all timelike $w^a$.

As another example, let us consider the case of dust ($P=0$ perfect
fluids). Of course, this case can be deduced from
the previous one by setting $P=0$. However, in dimension $N=5$
some stronger results can be derived. To see it, recall that
any 2-form $F_{[2]}$ with no null eigenvector can only exist in {\it odd}
dimension $N=2n+1$, and must take the form
$$
F_{[2]}=\mu_{1}(\bm{e}_1\wedge \bm{e}_2)+\dots +
\mu_{n}(\bm{e}_{2n-1}\wedge \bm{e}_{2n})
$$
where $\left\{\bm{e}_{0},\bm{e}_{1},...,\bm{e}_{2n}\right\}$ is an orthonormal
basis and $\mu_{i}$ ($i=1,\dots ,n$) are non-zero constants.
Thus, in the particular case that all the $\mu_{i}$'s are equal we
get for the superenergy tensor (\ref{se2}) of such an $F_{[2]}$
\be
T_{ab}\{F_{[2]}\}=\frac{\mu_{1}^2}{2}\left[n\, \bm{e}_0\otimes \bm{e}_0+
(2-n)(\bm{e}_1\otimes \bm{e}_1 +\dots +\bm{e}_{2n}\otimes
\bm{e}_{2n})\right]_{ab}\, .\label{2n+1}
\ee
\begin{coro}
In 5 dimensions, $T_{ab}$ is {\em algebraically}
the energy-momentum tensor of a dust, that is $T_{ab}=\rho u_{a} u_{b}$
where $(u\cdot u)=1$ and $\rho\geq 0$, if and only if $T_{ab}$ is
the s-e tensor of a 2-form $F_{[2]}$ with no null eigenvector having
$\mu_{2}=\mu_{1}$.
\end{coro}
\P
This is the case $n=2$ ($\Longrightarrow N=5$) of the previous
formula (\ref{2n+1}), identifying $\bm{u}=\bm{e}_{0}$ and $\rho =\mu_{1}^2$.
Notice that the timelike direction $\bm{u}$ is intrinsically defined by
$*(F_{[2]}\wedge  F_{[2]})$.\N

\subsection*{Acknowledgments}

The authors are grateful to the Ministerio de Educaci\'on, Cultura y Deporte
in Spain for support for a longer stay at the Department of Theoretical Physics
at the University of the Basque Country (SAB1999-0135 for G.B.) and
for a 3-month stay at the Institut f\"ur Theoretische Physik of the
University of Vienna (PR2000-0222 for J.M.M.S.). JMMS is thankful to
this Institut, where the paper was partly written, for kind hospitality.
We thank Anders H\"{o}glund for useful comments and Alfonso
Garc\'{\i}a-Parrado for pointing out some typos.

\end{document}